\documentclass[
twocolumn,
aps,
prl,
]{revtex4-2}
\usepackage{color}
\usepackage{hyperref}
\usepackage{amsmath, amssymb}
\hypersetup{colorlinks=true,breaklinks,linkcolor=blue,urlcolor=blue,citecolor=blue}
\usepackage{graphicx}% Include figure files
\usepackage{dcolumn}% Align table columns on decimal point
\usepackage{bm}% bold math
\usepackage{xspace}
\usepackage{ulem}

\def\vec#1{\boldsymbol #1}

\usepackage{ulem}
\usepackage{xcolor}
\DeclareRobustCommand{\erase}{\bgroup\markoverwith{\textcolor{red}{\rule[.5ex]{2pt}{0.4pt}}}\ULon}

\newcommand{\sP}{\mathcal{P}}

\newcommand{\ra}{\rangle}
\newcommand{\Ns}{N_{\text{s}}}

\usepackage{natbib}

\begin{document}
\preprint{}

\title{Comprehensive $ab$ $initio$ investigation of the phase diagram of quasi-one-dimensional molecular solids} 

\author{Kazuyoshi Yoshimi$^{1, \dagger}$, Takahiro Misawa$^{1, 2, \dagger}$, Takao Tsumuraya$^{3,4}$, and Hitoshi Seo$^{5, 6}$}
\affiliation{$^1$Institute for Solid State Physics,~University of Tokyo,~5-1-5 Kashiwanoha, Kashiwa, Chiba 277-8581, Japan}
\affiliation{$^2$Beijing Academy of Quantum Information Sciences, Haidian District, Beijing 100193, China}
\affiliation{$^3$Priority Organization for Innovation and Excellence, Kumamoto University, 2-39-1 Kurokami, Kumamoto 860-8555, Japan} 
\affiliation{$^4$Magnesium Research Center, Kumamoto University, 2-39-1 Kurokami, Kumamoto 860-8555, Japan}
\affiliation{$^5$Condensed Matter Theory Laboratory, RIKEN, Wako, Saitama 351-0198, Japan}
\affiliation{$^6$Center for Emergent Matter Science (CEMS), RIKEN, Wako, Saitama 351-0198, Japan}
\affiliation{\rm $^{\dagger}$The authors contributed to this work equally.}
\date{\today}% It is always \today, today,
             %  but any date may be explicitly specified

\begin{abstract}
An {\it ab initio} investigation of the family of molecular compounds TM$_2${\it X} is conducted, where TM is either TMTSF or TMTTF and {\it X} takes centrosymmetric monovalent anions. 
By deriving the extended Hubbard-type Hamiltonians from first-principles band calculations 
and evaluating not only the intermolecular transfer integrals but also the Coulomb parameters, we discuss their material dependence in the unified phase diagram. 
Furthermore, we apply the many-variable variational Monte Carlo method to accurately determine the symmetry-breaking phase transitions, and show the development of the charge and spin orderings. 
We show that the material-dependent parameter can be taken as the correlation effect, represented by the value of the screened on-site Coulomb interaction $U$ relative to the intrachain transfer integrals, for the comprehensive understanding of the spin and charge ordering in this system. 
\end{abstract}

\maketitle

{\it Introduction---.}~Molecular solids serve as textbook materials to study many-body physics in condensed matter, despite their complex crystal structures with many atoms in the unit cell, owing to the success in modeling the low-energy electronic properties based on the frontier molecular orbitals~\cite{Ishiguro_book,Lebed_book}. 
A general challenge is to understand multiple symmetry-broken phases observed in different systems in a systematic way, where the charge, spin, and lattice degrees of freedom are coupled to each other, based on the effective Hamiltonians and identifying the factors governing them~\cite{Seo_ChemRev}. 

The family of compounds constituted by TMTSF (= tetramethyltetraselenafulvalene) or TMTTF (= tetramethyltetrathiafulvalene) molecules, which we call together TM here, forming 2:1 salts with monovalent anions {\it X}$^{-1}$, TM$_2${\it X}, has been studied for decades as a typical example showing such a rich variety of phases~\cite{Jerome_ChemRev,Seo_ChemRev,Lebed_book,Brown_inbook}. 
TMTSF$_2${\it X} historically outstands as the first organic system undergoing superconductivity~\cite{Jerome_SC}. 
Unconventional superconductivity with an anisotropic gap function is realized, whose `glue' for the Cooper pair formation is the strong spin fluctuation, supported by the pressure-temperature phase diagram with the superconducting phase neighboring an antiferromagnetic (AFM) phase driven by the nesting of the Fermi surface~\cite{BealMonod_1986,Hasegawa_1987,Takigawa_1987,Shimahara_1989,Brown_review}. 

Later, the isostructural TMTTF counterparts became realized as strongly-correlated electron systems, owing to the atomic substitution resulting in smaller bandwidths than in the TMTSF compounds~\cite{Chow_2000,Brown_inbook,Kito_2021}. 
In particular, at high temperatures of the order of 100~K, TMTTF$_2${\it X} hosts a competition between a Mott insulating state, the so-called dimer-Mott (DM) insulator, and a charge ordering (CO) state with electron-rich and poor molecules spontaneously arranging to avoid the Coulomb repulsion: a common framework for the understanding of correlated quarter-filled  systems~\cite{Seo_1997,Seo_2000,Powell_review2006}. 
In the low-temperature region ($\sim10$K), on the other hand, where the magnetic interactions start to develop between the localized spins in the strongly correlated regime, the AFM state competes with the spin-Peierls instability accompanying lattice distortion owing to the quasi-one-dimensionality~\cite{Zamborszky_2002,Iwase_2011,Dressel_2012}. 

\begin{figure*}[t] 
\begin{center} 
\includegraphics[width=0.9 \textwidth]{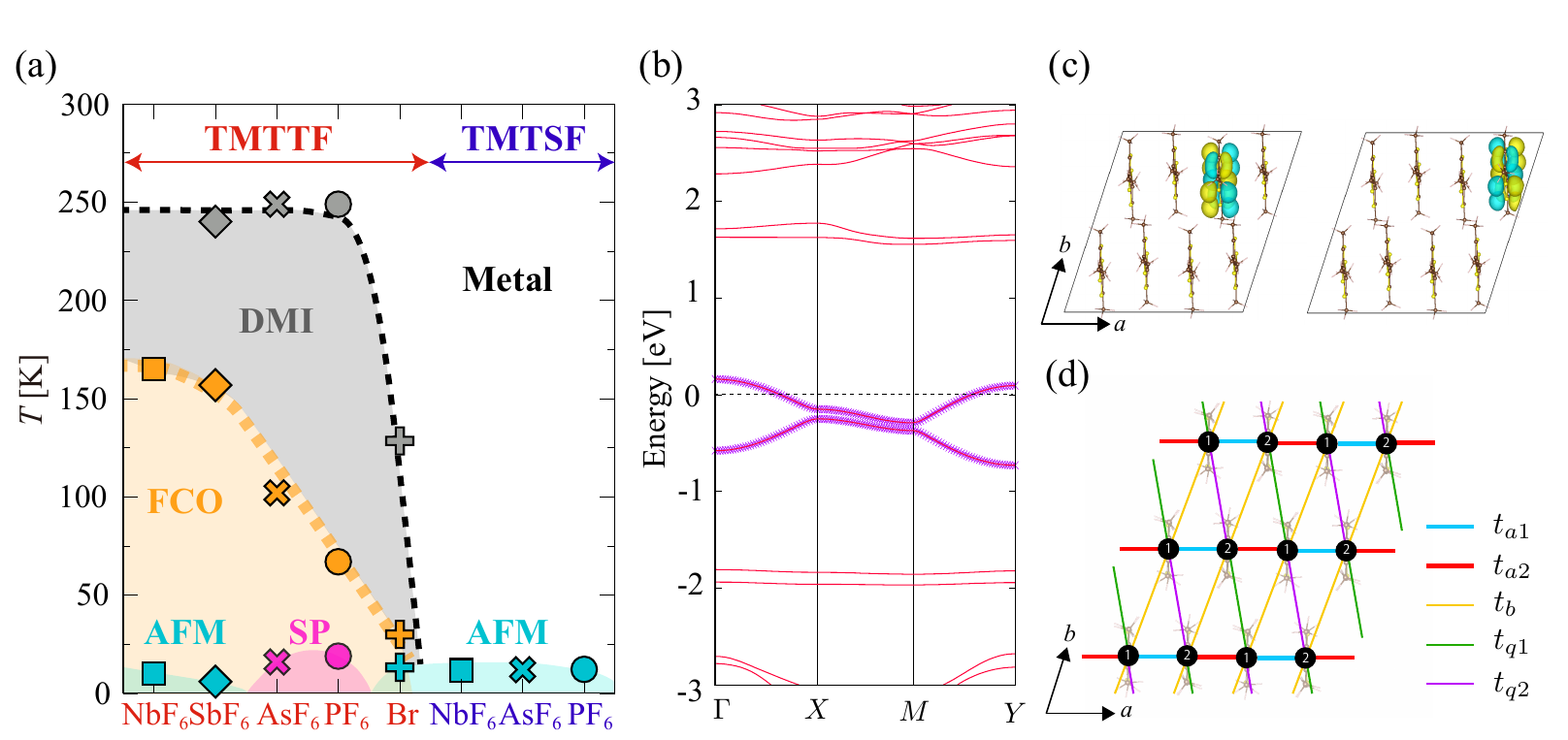}
\vspace{-0.5cm} 
\caption{
(a) Schematic phase diagram of (TM)$_2${\it X}. DMI, FCO, AFM, and SP indicate dimer-mott insulating, ferroelectric charge ordered, antiferromagnetic, and spin-Peierls phases, respectively. Symbols are the ambient-pressure transition temperatures for each material~\cite{supplement}. We note that, since the reported AFM transition temperatures for the TMTSF salts are the same~\cite{supplement}, the order is determined by the volume of the unit cell.
(b) Band structure of TMTTF$_2$PF$_6$. 
The thin solid lines are the DFT results, while the bold lines are obtained using the MLWFs, overlapping the valence bands. 
We set the Fermi energy to zero (the dotted line). 
(c) Drawings of the two MLWFs for TMTTF$_2$PF$_6$. 
(d) Definition of the transfer integrals. 
The TMTTF molecules and MLWFs are drawn by using \texttt{VESTA}~\cite{VESTA}.} 
\label{fig-bands}
\end{center}
\end{figure*}

The variation among the symmetry-breaking states, in which both the charge and spin degrees of freedom are involved,  depending on the applied pressure to the compounds and on the choice of the combination of TM and {\it X} (chemical pressure), has been extensively discussed~\cite{Kuwabara_2003,Sugiura_2005,Yoshimi_2012,Clay_review,Yoshioka_crystals_review}.
Nevertheless, there is room toward a comprehensive understanding of them. 
There are different views on how the choice of {\it X} and the amount of applied pressure affect the microscopic parameters, and the impact of the difference between TMTSF and TMTTF is still obscure. 
Conventionally, the two systems are drawn in a unified pressure-temperature phase diagram~\cite{Jerome_ChemRev,Brown_inbook,Yoshimi_2012,Clay_review}, as shown in Fig.~\ref{fig-bands}(a). 
Although there are different versions in the literature, 
here we simply draw the `ambient pressure' positions of the representative compounds 
in an equally spaced manner, 
together with the reported transition temperatures~\cite{PhysRevB.61.511,Sakata2006,cryst10121085,PhysRevB.84.035124,Kito_2021}. 

From the early days, in the TMTSF salts, it is discussed that the applied pressure increases the dimensionality~\cite{Jerome_SC}. 
On the other hand, the degree of dimerization along the one-dimensional chain direction was also pointed out to be an important factor, especially for the TMTTF salts~\cite{Nogami2005}. 
And more recently, to understand the complicated interplay between the phases toward the left-hand end of the phase diagram, it was proposed that the CO instability can be the controlling factor~\cite{Yoshimi_2012}; this suggests the essential role of electron correlation. 
Whether we can choose a parameter that varies monotonically as we go along the compounds listed in the horizontal axis is the question we ask here. 

In this work, to investigate these issues,  we apply first-principles density functional theory (DFT) calculations~\cite{H-K_1964, Kohn_Sham} to a number of members of the TM$_2${\it X} family with centrosymmetric $X$; we numerically derive not only the transfer integrals between the molecules but also the correlation parameters. 
We then use these parameters as inputs to a highly-accurate numerical solver, i.e., the many-variable variational Monte Carlo (mVMC) method~\cite{Tahara_JPSJ2008,misawa_CPC2019,mVMC}, for the effective extended Hubbard-type Hamiltonian and seek possible long-range ordering in the charge and spin degree of freedoms. 
As a result, we can directly compare with the experimentally observed phases and comprehensively extract the determining 
factor at a quantitative level. 

{\it Ab initio derivation of microscopic parameters---.}~The DFT calculations are performed using \texttt{Quantum Espresso (version 6.6)}~\cite{QE} for the experimental crystal structures~\cite{Kito_2021}. 
We employ norm-conserving pseudopotentials based on the Vanderbilt formalism with plane-wave basis sets~\cite{Hamann_ONCV2013, Schlipf_CPC2015}. 
The exchange-correlation functional used in this study is the generalized gradient approximation by Perdew, Burke, and Ernzerhof~\cite{GGA_PBE}, which is commonly used for accurately describing electronic states of molecular compounds~\cite{Ishibashi-2009}.
The cutoff energies for plane waves and charge densities are $70$ and $280$ Ry, respectively. A $7\times 7\times 3$ uniform $\bm{k}$-point mesh was used with a Gaussian smearing method during self-consistent loops.

For deriving the interaction parameters, we use the constrained random phase approximation (cRPA) method~\cite{PhysRevB.70.195104, Imada_JPSJ2010}. 
In fact, previous studies applying the cRPA method to molecular solids show good agreement between the results for the derived effective Hamiltonians and the experiments ~\cite{Shinaoka2012, PhysRevResearch.2.032072, PhysRevResearch.3.043224, Ido2022, PhysRevB.105.205123, PhysRevB.102.235116}.
On the basis of the obtained DFT electronic states, we construct maximally localized Wannier function(MLWF)s and derive the parameters using \texttt{RESPACK}~\cite{RESPACK}. In the calculations, the energy cutoff for the dielectric function was set to be $3$ Ry. 

Figures~\ref{fig-bands}(b) and (c) show the band structure and the MLWFs, respectively, of TMTTF$_2$PF$_6$ as an example. 
The two bands crossing the Fermi energy are constituted from the bonding and anti-bonding combination of the highest occupied molecular orbitals situated on the two TMTTF molecules in the unit cell shown in Fig.~\ref{fig-bands}(b). 
Using these MLWFs, we evaluated the transfer integrals and  the density-density interactions, i.e. the on-site and off-site Coulomb interactions (the exchange interactions are negligibly small so we omit them in the following). 
The obtained microscopic parameters are listed in Table II in the supplemental materials~\cite{supplement}.
We adopt the indices of bonds from refs.~\cite{Yoshimi_2012,Yoshimi2012_PhysicaB}, as shown in Fig. \ref{fig-bands}(d).

Then the following {\it ab initio} extended Hubbard-type Hamiltonian is obtained:
\begin{align}
&H=
\sum_{ij,\sigma}t_{ij}(c_{i\sigma}^{\dagger}c_{j\sigma}+{\rm h.c.})
+U\sum_{i}n_{i\uparrow}n_{i\downarrow}
+\sum_{ij}V_{ij}N_{i}N_{j}
\label{Ham}
\end{align}
where $c^{\dagger}_{i\sigma}$ and $c_{i\sigma}$ are the creation and annihilation operators of an electron with spin $\sigma$ at the $i$-th site, respectively.
The number operators are defined as $n_{i\sigma}=c_{i\sigma}^{\dagger}c_{i\sigma}$ and
$N_{i}=n_{i\uparrow}+n_{i\downarrow}$.$t_{ij}$, $U$, and $V_{ij}$ represent the transfer integrals, on-site and off-site Coulomb interactions, respectively.
We note that, although TM$_2${\it X} show quasi-one-dimensional electronic band structures, the Coulomb parameters show large values also for the interchain bonds. 

\begin{figure}[t] 
\begin{center} 
\includegraphics[width=0.45 \textwidth]{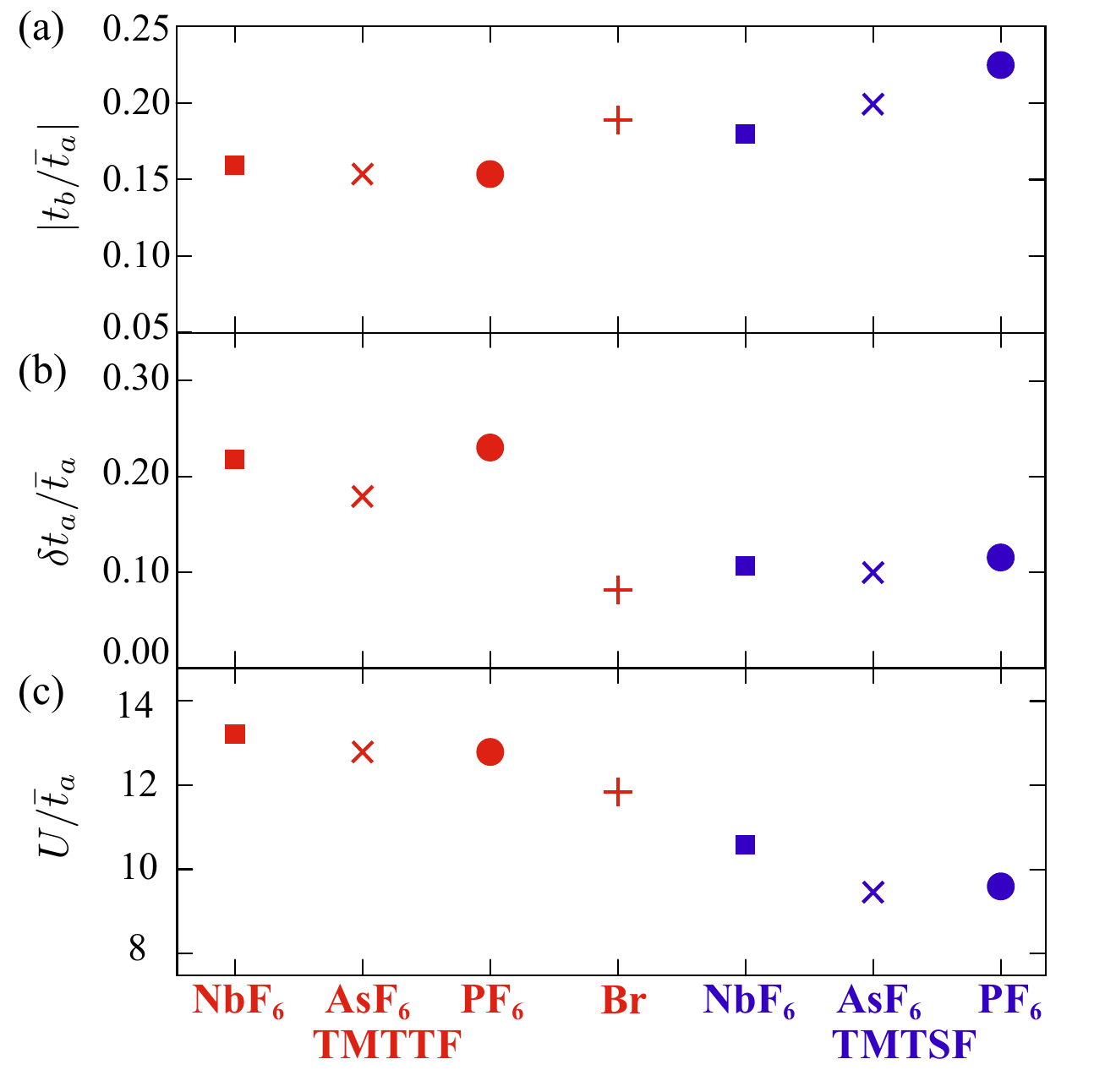}
\vspace{-0.5cm}
\caption{
(a) Transfer integral between TM molecules along $b$-axis ($t_b$), (b) difference in the alternating transfer integrals along $a$-axis ($\delta t_a \equiv |t_{a2}-t_{a1}|$), and (c) the on-site Coulomb interaction ($U$) for TM$_2${\it X} (TM = TMTSF, TMTTF, {\it X} = NbF$_6$, AsF$_6$, Br, and PF$_6$). Each value is normalized by the mean $a$-axis transfer integral, $\bar{t}_a\equiv |(t_{a1}+t_{a2})/2|$.} 
\label{fig-model}
\end{center}
\end{figure}

First, let us discuss the material dependence of intermolecular transfer energies based on our evaluation.
Since it is known that the temperature variation of structural parameters is large~\cite{Jacko2013},  here we show the results for TMTTF$_2${\it X} with {\it X}= NbF$_6$, AsF$_6$, Br, and PF$_6$ at fixed $200$~K, and for TMTSF$_2${\it X} with {\it X}= NbF$_6$, AsF$_6$, and PF$_6$ at room temperature~\cite{supplement}.
In Fig.~\ref{fig-model}, we show the variation of three different values along the horizontal axis of the phase diagram Fig.~\ref{fig-bands}(a). 
The first two, the degree of interchain couplings and the degree of dimerization plotted in Figs.~\ref{fig-model}(a) and \ref{fig-model}(b), respectively, have been discussed in the literature as mentioned above. 
We renormalize them by the mean transfer integral along the $a$-axis, $\bar{t}_a\equiv|(t_{a1}+t_{a2})/2|$, which is larger than those along the other directions. 

The largest interchain transfer integral, which is $t_b$, is plotted in Fig.~\ref{fig-model}(a); the interlayer transfer integrals along the $c$-axis are negligible. 
Our results show larger $|t_b/\bar{t}_{a}|$ for the TMTSF salts than the TMTTF salts as a general trend;  
TMTSF has a broader extent of the wave function in the transverse direction than TMTTF, owing to the difference between the Se and S atoms at the sides of the molecules. 
As for the degree of dimerization, $\delta t_a / \bar{t}_{a} \equiv |t_{a2}-t_{a1}| / \bar{t}_{a}$ [Fig.~\ref{fig-model}(b)],  the overall tendency is that it decreases from the left to the right side, while some deviates from the monotonic variation; for example, TMTTF$_2$Br shows a similar value to the TMTSF salts. 

Finally, Fig.~\ref{fig-model}~(c) shows the on-site Coulomb interaction, $U$, which is a measure of the strength of electron correlation. 
The intermolecular parameters, $V_{ij}$, also show similar tendencies. 
We can see that $U/\bar{t}_a$ monotonically decrease from the left to the right side, except for the two salts at the right end: TMTSF$_2$PF$_6$ ($U/\bar{t}_a=9.59$) is slightly larger than TMTSF$_2$AsF$_6$ ($U/\bar{t}_a=9.46$). 
Since the electron correlation is responsible for all the phases, from the strongly correlated DM insulating and CO phases, to the nesting-driven AFM and the spin-fluctuation mediated superconducting phases, this behavior is consistent with the temperature scale for their occurrence, decreasing from left to right of the phase diagram in Fig.~\ref{fig-bands}(a). 

This tendency in $U/\bar{t}_{a}$ can be understood by the effect of chemical pressure. More concretely, first, by the unit cell volume decreasing from left to right, the transfer integrals reflecting the overlap between MLWFs increase. 
Next, since the screening effect becomes stronger with increasing the band width, the screened $U$ is smaller with increasing the pressure. As a combination of them, together with the difference in the bare onsite Coulomb parameter $U_{\rm bare}$ between the molecules, $U/\bar{t}_a$ monotonically decreases from left to right in Fig.~\ref{fig-model} (c). For more detailed information about the parameters and their numerical accuracy, see ref.~\cite{supplement}.

{\it mVMC analysis---.}~To directly detect the symmetry-breaking phase transitions, we solve the $ab$ $initio$ effective Hamiltonian by using the mVMC method~\cite{Tahara_JPSJ2008,misawa_CPC2019,mVMC}.
The form of the variational wave function used in this study is given by
\begin{eqnarray}
|\psi\ra =\mathcal{L}_{S}\sP_{\rm G}\sP_{\rm J}|\phi_{\rm pair}\ra,
\label{mvmc_wavefunction_def}
\end{eqnarray}
where $\sP_{\rm G}$ ($\sP_{\rm J}$) represents the Gutzwiller factor~\cite{Gutzwiller_PRL1963} (Jastrow factor~\cite{Jastrow_PR1955}) and $\mathcal{L}_{S}$ is the total spin projection.
We use the spin singlet total spin projection ($S=0$) for the ground states.
The pair-product wave function $|\phi_{\rm pair}\ra$ is defined as
\begin{eqnarray}
|\phi_{\rm pair}\ra= \Big[{\sum_{i,j=1}^{\Ns}
}f_{ij}{c_{i\uparrow}^{\dag}c_{j\downarrow}^{\dag}}\Big]^{N_{e}/2} |0 \ra,
\end{eqnarray}
where $f_{ij}$ represents the variational parameters, and
$\Ns$ and $N_e$ are the number of sites and electrons, respectively. 
In the mVMC calculations, we map the lattice structure shown in Fig.~\ref{fig-bands}(d) onto the square lattice without loss of generality.
The total number of sites is given by $\Ns=L_{x}\times L_{y}$.
We impose a $4\times2$ sublattice structure and the periodic boundary conditions, and optimize all the variational parameters simultaneously 
using the stochastic reconfiguration method~\cite{Sorella_PRB2001,Tahara_JPSJ2008}
to obtain the ground state. 

We consider all the bonds that are listed in Table~II in the supplemental materials~\cite{supplement};  they contain all the bonds shown in Fig.~\ref{fig-bands}(d) as well as $V_{2a}$ that is the next-nearest-neighbor Coulomb interaction along the chain direction. 
We note that the effective Coulomb interactions are reduced by a constant shift $\Delta_{\rm DDF}=0.20$ eV~\cite{supplement}, adopting the results of the previous study~\cite{Nakamura_JPSJ2010, PhysRevB.86.205117}, to take into account dimensional screening effects. 
We calculate the spin/charge structure factors defined as
\begin{align}
N(\vec{q})&=\frac{1}{N_{\rm s}}\sum_{i,j}\langle N_{i}\cdot N_{j}\rangle e^{i\vec{q}(\vec{r}_{i}-\vec{r}_{j})},\\
S(\vec{q})&=\frac{1}{N_{\rm s}}\sum_{i,j}\langle \vec{S}_{i}\cdot \vec{S}_{j}\rangle e^{i\vec{q}(\vec{r}_{i}-\vec{r}_{j})},
\end{align}
where $N_{i}$ ($\vec{S_{i}}$) represents the number (spin) operator at site $i$.
The Fourier transformation is performed for the mapped square lattice.

\begin{figure}[t] 
\begin{center} 
\includegraphics[width=8cm]{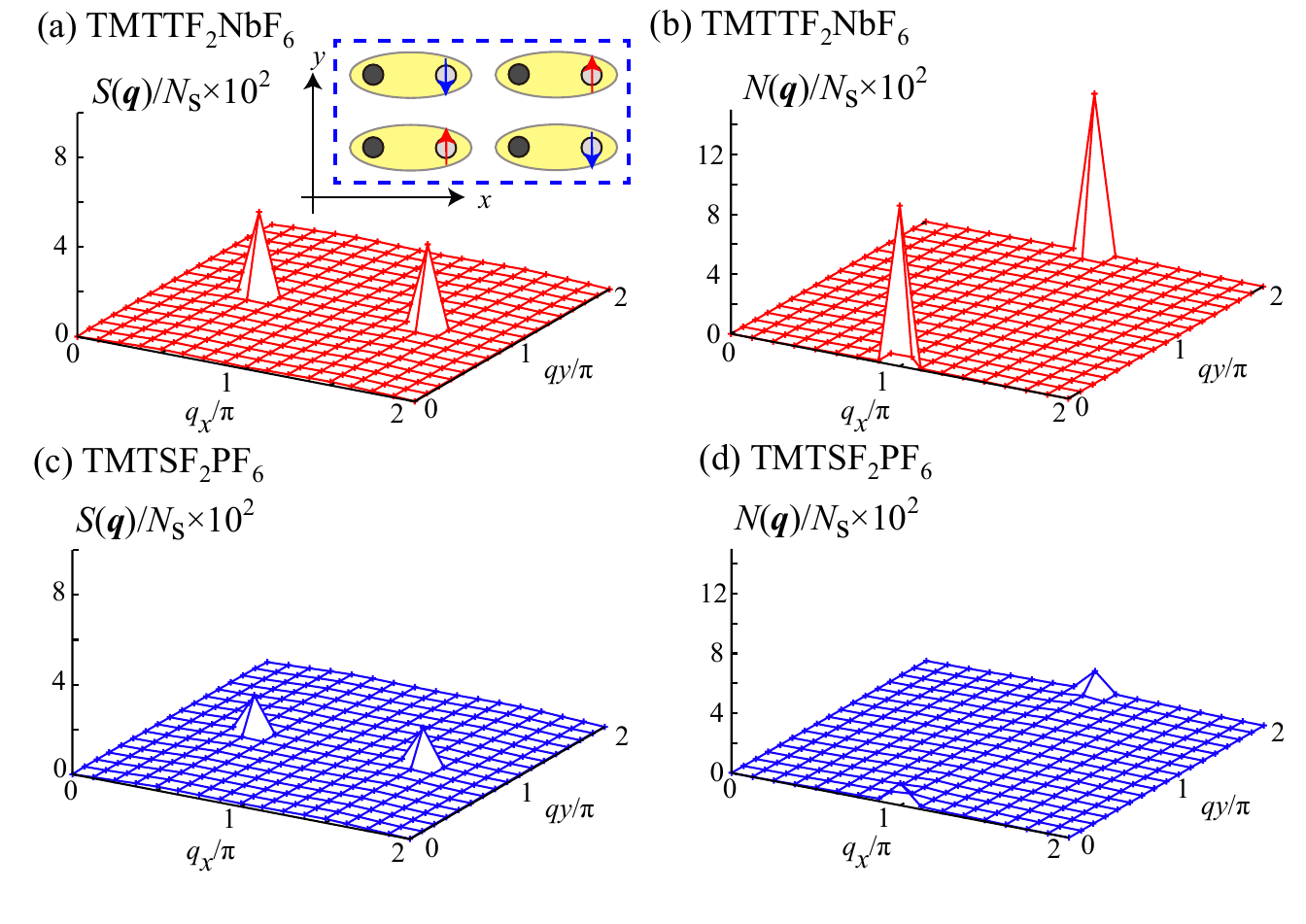}
\caption{
Spin [$S(\vec{q})$] and charge [$N(\vec{q})$] structure factors for (a), (b) TMTTF$_{2}$NbF$_{6}$ and (c), (d) TMTSF$_{2}$PF$_{6}$ for $L_{x}=L_{y}=16$.
In the inset, we show the schematic spin and charge-ordered pattern of TMTTF$_{2}$NbF$_{6}$ in the real space, where the yellow oval represents the dimer.
The broken blue line represents the $4\times2$ sublattice structure used in mVMC calculations.}
\label{fig:Sq}
\end{center}
\end{figure}

Let us present $S(\vec{q})$ and $N(\vec{q})$ for the two end members of the phase diagram,  TMTTF$_2$NbF$_{6}$ and TMTSF$_2$PF$_{6}$. 
As shown in Figs.~\ref{fig:Sq}(a) and \ref{fig:Sq}(b), for TMTTF$_2$NbF$_{6}$, which has the strongest electronic correlation, the spin (charge) structure factors has a sharp peak at $\vec{q}_{\rm peak}^s=(\pi/2,\pi)$ [$\vec{q}_{\rm peak}^c=(\pi,0)$].
The corresponding ordering pattern in the real space is shown in the inset of Fig.~\ref{fig:Sq}(a).
Since these peak values remain finite after size extrapolation, 
the CO state with spin ordering is the ground state, 
consistent with previous studies~\cite{Seo_1997,Yoshimi_2012}.
Note that, in the previous studies the long range Coulomb interactions are introduced by hand
to stabilize the CO state with 
the ferroelectric pattern which is observed experimentally. 
We emphasize that, here 
the condition favoring 
the ferroelectric CO state is satisfied from the $ab$ $initio$ way, displayed in  Fig. S1 (d) in the supplemental materials~\cite{supplement}.
In contrast, we find that the peak values of the structure factors for parameters of TMTSF$_2$PF$_{6}$ become significantly small as shown in Figs.~\ref{fig:Sq}(c) and \ref{fig:Sq}(d).
This result is consistent with our results above that TMTSF$_2$PF$_{6}$ has much weaker electronic correlation.

\begin{figure}[t] 
\begin{center} 
\includegraphics[width=0.5 \textwidth]{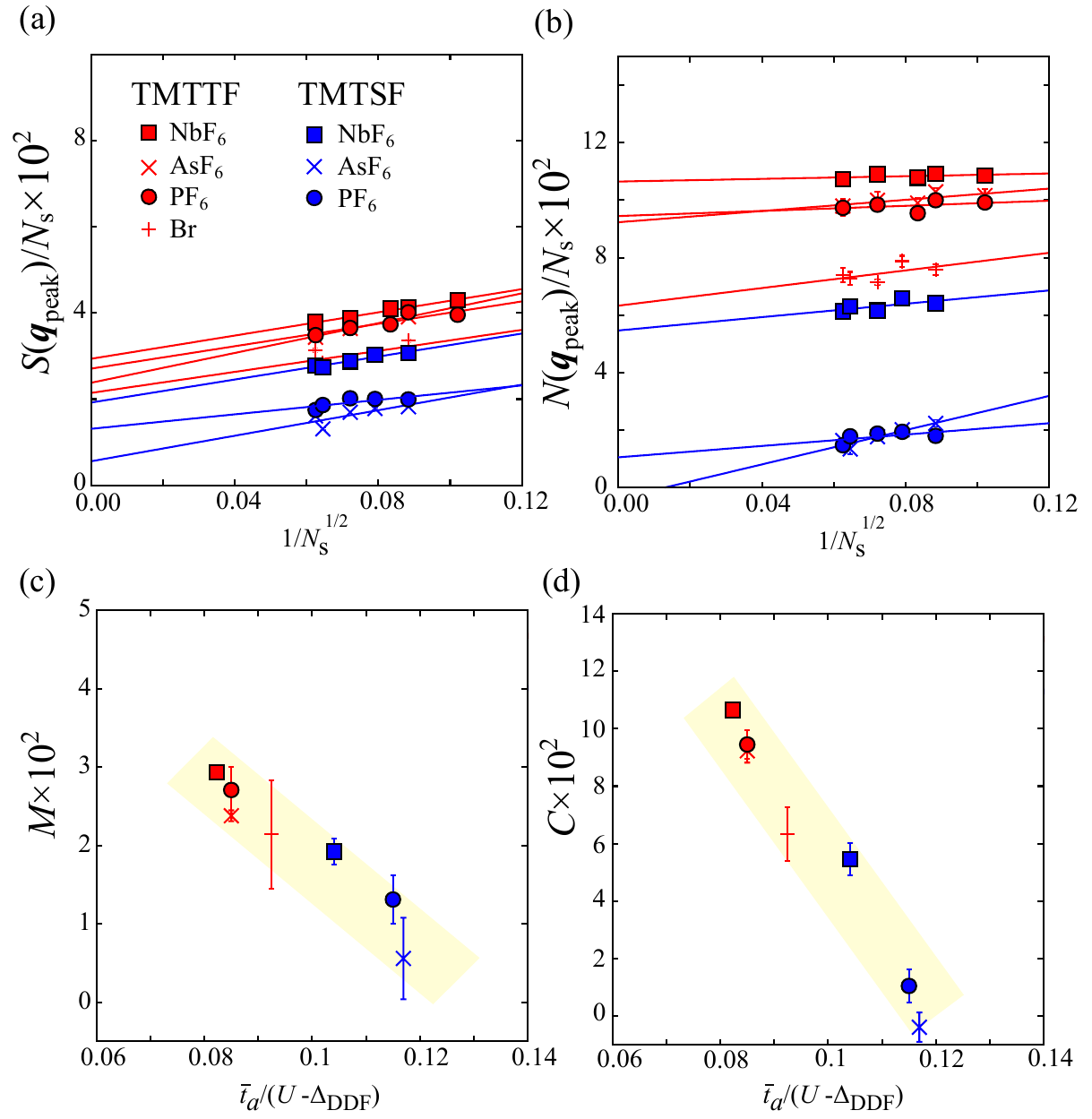}
\caption{
Size dependence of (a) the spin and (b) the charge
structure factors for 5 TMTTF salts and 3 TMTSF salts.
Coulomb interaction dependence of the
order parameter of (c) the spin ordering and (d) the charge ordering.
Both the charge and spin order gradually decrease as
a function of the Coulomb interaction. 
}
\label{fig:NqSq}
\end{center}
\end{figure}

To clarify the material dependence, we perform the mVMC calculations for all the members and conduct the size extrapolation for the obtained peak values, in the spin and charge structure factors, as shown in Figs.~\ref{fig:NqSq}(a) and (b), respectively.  
The thermodynamic limits are defined as $M=\lim_{N_{\rm s}\rightarrow\infty}S(\vec{q}_{\rm peak}^s)/N_{\rm s}$,
$C=\lim_{N_{\rm s}\rightarrow\infty}N(\vec{q}_{\rm peak}^c)/N_{\rm s}$.
Both the spin and charge structure factors show nearly linear dependencies as a function of the linear dimension of the number of sites, i.e., $N_{\rm s}^{1/2}$.

The material dependence of $M$ and $C$ is shown in Figs.~\ref{fig:NqSq} (c) and (d), respectively, as a function of the inverse of the normalized Coulomb interaction, $(U-\Delta_{\rm DDF})/\bar{t}_a$. 
We find that the amplitude of the charge and spin orders decrease monotonically with respect to the amplitude of the Coulomb interactions.
This tendency is consistent with the experimental results where the transition temperature of CO decreases by varying the material from left to right.
At the border, TMTSF$_2$AsF$_{6}$ and TMTSF$_2$PF$_{6}$ show a small but finite value of $M$ of the order of 10$^{-2}$, and $C$ almost vanishes. 

These results, combined with the {\it ab initio} results in Fig.~\ref{fig-model}(c), both clearly show that the charge and spin orderings are governed by the relative amplitude of the Coulomb interactions.
This is the first $ab$ $initio$ evidence that the correlation effects play key roles in stabilizing the spin/charge ordering in TM$_2X$ and serve as the good index for the horizontal axis of the unified phase diagram. The success here also indicates that our low-energy effective model is reliable at a quantitative level, considering the high precision of the mVMC method. This is owing to the fact that the band structure near the Fermi energy is simple, as shown in Fig.~\ref{fig-bands} (b), which guarantees the accuracy of the cRPA method, since it incorporates the effect of bands other than the target bands. In fact, in many of the molecular solids, bands near the Fermi energy are isolated, thus it is expected that our approach can be applicable to a wide range of functional molecular materials.

{\it Perspectives---.}~We have shown that the material dependence in the unified phase diagram for TM$_2X$ is governed by the correlation effect, represented by the on-site Coulomb interaction $U$, which gives a comprehensive understanding of the spin and charge orderings in this system. 
One point that disagrees with the experiments is that in our results, spin patterns with long periodicity such as incommensurate AFM (spin-density-wave) states suggested experimentally 
are difficult to reproduce in our finite-size systems. 
Finally, superconductivity is not found in our study, which is consistent with experiments showing it under pressure for the compounds we studied; its {\it ab initio} investigation is left for future studies. 

\begin{acknowledgements}
We thank R. Kato, S. Kitou, and T. Nakamura for fruitful discussions. This work was supported by MEXT/JSPJ KAKENHI under grant numbers 19K03723, 19K21860, 20H04463, 21H01041, 22K03526, 23H03818, and 23H04047. 
KY and TM were supported by Building of Consortia for the Development of Human Resources in Science and Technology, MEXT, Japan.
The computation in this work was performed using the 
facilities of the Supercomputer Center, Institute for Solid State Physics, University of Tokyo.
\end{acknowledgements}
\bibliography{main}

%apsrev4-2.bst 2019-01-14 (MD) hand-edited version of apsrev4-1.bst
%Control: key (0)
%Control: author (8) initials jnrlst
%Control: editor formatted (1) identically to author
%Control: production of article title (0) allowed
%Control: page (0) single
%Control: year (1) truncated
%Control: production of eprint (0) enabled
\providecommand{\noopsort}[1]{}\providecommand{\singleletter}[1]{#1}%
\begin{thebibliography}{61}%
\makeatletter
\providecommand \@ifxundefined [1]{%
 \@ifx{#1\undefined}
}%
\providecommand \@ifnum [1]{%
 \ifnum #1\expandafter \@firstoftwo
 \else \expandafter \@secondoftwo
 \fi
}%
\providecommand \@ifx [1]{%
 \ifx #1\expandafter \@firstoftwo
 \else \expandafter \@secondoftwo
 \fi
}%
\providecommand \natexlab [1]{#1}%
\providecommand \enquote  [1]{``#1''}%
\providecommand \bibnamefont  [1]{#1}%
\providecommand \bibfnamefont [1]{#1}%
\providecommand \citenamefont [1]{#1}%
\providecommand \href@noop [0]{\@secondoftwo}%
\providecommand \href [0]{\begingroup \@sanitize@url \@href}%
\providecommand \@href[1]{\@@startlink{#1}\@@href}%
\providecommand \@@href[1]{\endgroup#1\@@endlink}%
\providecommand \@sanitize@url [0]{\catcode `\\12\catcode `\$12\catcode
  `\&12\catcode `\#12\catcode `\^12\catcode `\_12\catcode `\%12\relax}%
\providecommand \@@startlink[1]{}%
\providecommand \@@endlink[0]{}%
\providecommand \url  [0]{\begingroup\@sanitize@url \@url }%
\providecommand \@url [1]{\endgroup\@href {#1}{\urlprefix }}%
\providecommand \urlprefix  [0]{URL }%
\providecommand \Eprint [0]{\href }%
\providecommand \doibase [0]{https://doi.org/}%
\providecommand \selectlanguage [0]{\@gobble}%
\providecommand \bibinfo  [0]{\@secondoftwo}%
\providecommand \bibfield  [0]{\@secondoftwo}%
\providecommand \translation [1]{[#1]}%
\providecommand \BibitemOpen [0]{}%
\providecommand \bibitemStop [0]{}%
\providecommand \bibitemNoStop [0]{.\EOS\space}%
\providecommand \EOS [0]{\spacefactor3000\relax}%
\providecommand \BibitemShut  [1]{\csname bibitem#1\endcsname}%
\let\auto@bib@innerbib\@empty
%</preamble>
\bibitem [{\citenamefont {Ishiguro}\ \emph {et~al.}(1998)\citenamefont
  {Ishiguro}, \citenamefont {Yamaji},\ and\ \citenamefont
  {Saito}}]{Ishiguro_book}%
  \BibitemOpen
  \bibfield  {author} {\bibinfo {author} {\bibfnamefont {T.}~\bibnamefont
  {Ishiguro}}, \bibinfo {author} {\bibfnamefont {K.}~\bibnamefont {Yamaji}},\
  and\ \bibinfo {author} {\bibfnamefont {G.}~\bibnamefont {Saito}},\ }\href
  {https://doi.org/10.1007/978-3-642-58262-2} {\emph {\bibinfo {title} {Organic
  superconductors}}},\ \bibinfo {edition} {2nd}\ ed.,\ Springer series in
  solid-state sciences, 88\ (\bibinfo  {publisher} {Springer},\ \bibinfo
  {address} {Berlin; New York},\ \bibinfo {year} {1998})\BibitemShut {NoStop}%
\bibitem [{\citenamefont {Lebed}()}]{Lebed_book}%
  \BibitemOpen
  \bibfield  {author} {\bibinfo {author} {\bibfnamefont {A.}~\bibnamefont
  {Lebed}},\ }\bibfield  {title} {\bibinfo {title} {{The Physics of organic
  superconductors and conductors}},\ }\href@noop {} {\bibinfo  {journal}
  {Springer Series in Materials Science (Springer, Berlin, Heidelberg, 2008)}\
  }\BibitemShut {NoStop}%
\bibitem [{\citenamefont {Seo}\ \emph {et~al.}(2004)\citenamefont {Seo},
  \citenamefont {Hotta},\ and\ \citenamefont {Fukuyama}}]{Seo_ChemRev}%
  \BibitemOpen
\bibfield  {journal} {  }\bibfield  {author} {\bibinfo {author} {\bibfnamefont
  {H.}~\bibnamefont {Seo}}, \bibinfo {author} {\bibfnamefont {C.}~\bibnamefont
  {Hotta}},\ and\ \bibinfo {author} {\bibfnamefont {H.}~\bibnamefont
  {Fukuyama}},\ }\bibfield  {title} {\bibinfo {title} {Toward systematic
  understanding of diversity of electronic properties in low-dimensional
  molecular solids},\ }\href {https://doi.org/10.1021/cr030646k} {\bibfield
  {journal} {\bibinfo  {journal} {Chem. Rev.}\ }\textbf {\bibinfo {volume}
  {104}},\ \bibinfo {pages} {5005} (\bibinfo {year} {2004})}\BibitemShut
  {NoStop}%
\bibitem [{\citenamefont {J\'{e}rome}(2004)}]{Jerome_ChemRev}%
  \BibitemOpen
  \bibfield  {author} {\bibinfo {author} {\bibfnamefont {D.}~\bibnamefont
  {J\'{e}rome}},\ }\bibfield  {title} {\bibinfo {title} {{Organic Conductors:
  From Charge Density Wave TTF{-}TCNQ to Superconducting (TMTSF)$_2$PF$_6$}},\
  }\href {https://doi.org/10.1021/cr030652g} {\bibfield  {journal} {\bibinfo
  {journal} {Chem. Rev.}\ }\textbf {\bibinfo {volume} {104}},\ \bibinfo {pages}
  {5565} (\bibinfo {year} {2004})}\BibitemShut {NoStop}%
\bibitem [{\citenamefont {Brown}\ \emph {et~al.}(2008)\citenamefont {Brown},
  \citenamefont {Chaikin},\ and\ \citenamefont {Naughton}}]{Brown_inbook}%
  \BibitemOpen
  \bibfield  {author} {\bibinfo {author} {\bibfnamefont {S.~E.}\ \bibnamefont
  {Brown}}, \bibinfo {author} {\bibfnamefont {P.~M.}\ \bibnamefont {Chaikin}},\
  and\ \bibinfo {author} {\bibfnamefont {M.~J.}\ \bibnamefont {Naughton}},\
  }\bibinfo {title} {La tour des sels de bechgaard},\ in\ \href
  {https://doi.org/10.1007/978-3-540-76672-8_5} {\emph {\bibinfo {booktitle}
  {The Physics of Organic Superconductors and Conductors}}},\ \bibinfo {editor}
  {edited by\ \bibinfo {editor} {\bibfnamefont {A.}~\bibnamefont {Lebed}}}\
  (\bibinfo  {publisher} {Springer},\ \bibinfo {address} {Berlin, Heidelberg},\
  \bibinfo {year} {2008})\ pp.\ \bibinfo {pages} {49--87}\BibitemShut {NoStop}%
\bibitem [{\citenamefont {J\'erome}\ \emph {et~al.}(1980)\citenamefont
  {J\'erome}, \citenamefont {Mazaud}, \citenamefont {Ribault},\ and\
  \citenamefont {Bechgaard}}]{Jerome_SC}%
  \BibitemOpen
  \bibfield  {author} {\bibinfo {author} {\bibfnamefont {D.}~\bibnamefont
  {J\'erome}}, \bibinfo {author} {\bibfnamefont {A.}~\bibnamefont {Mazaud}},
  \bibinfo {author} {\bibfnamefont {M.}~\bibnamefont {Ribault}},\ and\ \bibinfo
  {author} {\bibfnamefont {K.}~\bibnamefont {Bechgaard}},\ }\bibfield  {title}
  {\bibinfo {title} {{Superconductivity in a synthetic organic conductor
  (TMTSF)$_2$PF$_6$}},\ }\href
  {https://doi.org/10.1051/jphyslet:0198000410409500} {\bibfield  {journal}
  {\bibinfo  {journal} {J. Physique Lett.}\ }\textbf {\bibinfo {volume} {41}},\
  \bibinfo {pages} {95} (\bibinfo {year} {1980})}\BibitemShut {NoStop}%
\bibitem [{\citenamefont {B\'eal-Monod}\ \emph {et~al.}(1986)\citenamefont
  {B\'eal-Monod}, \citenamefont {Bourbonnais},\ and\ \citenamefont
  {Emery}}]{BealMonod_1986}%
  \BibitemOpen
  \bibfield  {author} {\bibinfo {author} {\bibfnamefont {M.~T.}\ \bibnamefont
  {B\'eal-Monod}}, \bibinfo {author} {\bibfnamefont {C.}~\bibnamefont
  {Bourbonnais}},\ and\ \bibinfo {author} {\bibfnamefont {V.~J.}\ \bibnamefont
  {Emery}},\ }\bibfield  {title} {\bibinfo {title} {{Possible superconductivity
  in nearly antiferromagnetic itinerant fermion systems}},\ }\href
  {https://doi.org/10.1103/PhysRevB.34.7716} {\bibfield  {journal} {\bibinfo
  {journal} {Phys. Rev. B}\ }\textbf {\bibinfo {volume} {34}},\ \bibinfo
  {pages} {7716} (\bibinfo {year} {1986})}\BibitemShut {NoStop}%
\bibitem [{\citenamefont {Hasegawa}\ and\ \citenamefont
  {Fukuyama}(1987)}]{Hasegawa_1987}%
  \BibitemOpen
  \bibfield  {author} {\bibinfo {author} {\bibfnamefont {Y.}~\bibnamefont
  {Hasegawa}}\ and\ \bibinfo {author} {\bibfnamefont {H.}~\bibnamefont
  {Fukuyama}},\ }\bibfield  {title} {\bibinfo {title} {{NMR Relaxation Time of
  the Anisotropic Superconducting State in Quasi-One-Dimensional Systems}},\
  }\href {https://doi.org/10.1143/JPSJ.56.877} {\bibfield  {journal} {\bibinfo
  {journal} {J. Phys. Soc. Jpn.}\ }\textbf {\bibinfo {volume} {56}},\ \bibinfo
  {pages} {877} (\bibinfo {year} {1987})}\BibitemShut {NoStop}%
\bibitem [{\citenamefont {Takigawa}\ \emph {et~al.}(1987)\citenamefont
  {Takigawa}, \citenamefont {Yasuoka},\ and\ \citenamefont
  {Saito}}]{Takigawa_1987}%
  \BibitemOpen
  \bibfield  {author} {\bibinfo {author} {\bibfnamefont {M.}~\bibnamefont
  {Takigawa}}, \bibinfo {author} {\bibfnamefont {H.}~\bibnamefont {Yasuoka}},\
  and\ \bibinfo {author} {\bibfnamefont {G.}~\bibnamefont {Saito}},\ }\bibfield
   {title} {\bibinfo {title} {Proton spin relaxation in the superconducting
  state of (tmtsf)2clo4},\ }\href {https://doi.org/10.1143/JPSJ.56.873}
  {\bibfield  {journal} {\bibinfo  {journal} {J. Phys. Soc. Jpn}\ }\textbf
  {\bibinfo {volume} {56}},\ \bibinfo {pages} {873} (\bibinfo {year}
  {1987})}\BibitemShut {NoStop}%
\bibitem [{\citenamefont {Shimahara}(1989)}]{Shimahara_1989}%
  \BibitemOpen
  \bibfield  {author} {\bibinfo {author} {\bibfnamefont {H.}~\bibnamefont
  {Shimahara}},\ }\bibfield  {title} {\bibinfo {title} {{Long-Range
  Spin-Fluctuations and Superconductivity in Quasi-One-Dimensional Organic
  Compounds}},\ }\href {https://doi.org/10.1143/JPSJ.58.1735} {\bibfield
  {journal} {\bibinfo  {journal} {J. Phys. Soc. Jpn.}\ }\textbf {\bibinfo
  {volume} {58}},\ \bibinfo {pages} {1735} (\bibinfo {year}
  {1989})}\BibitemShut {NoStop}%
\bibitem [{\citenamefont {Brown}(2015)}]{Brown_review}%
  \BibitemOpen
  \bibfield  {author} {\bibinfo {author} {\bibfnamefont {S.~E.}\ \bibnamefont
  {Brown}},\ }\bibfield  {title} {\bibinfo {title} {{Organic superconductors:
  The Bechgaard salts and relatives}},\ }\href
  {https://doi.org/https://doi.org/10.1016/j.physc.2015.02.030} {\bibfield
  {journal} {\bibinfo  {journal} {Physica C}\ }\textbf {\bibinfo {volume}
  {514}},\ \bibinfo {pages} {279} (\bibinfo {year} {2015})}\BibitemShut
  {NoStop}%
\bibitem [{\citenamefont {Chow}\ \emph {et~al.}(2000)\citenamefont {Chow},
  \citenamefont {Zamborszky}, \citenamefont {Alavi}, \citenamefont {Tantillo},
  \citenamefont {Baur}, \citenamefont {Merlic},\ and\ \citenamefont
  {Brown}}]{Chow_2000}%
  \BibitemOpen
  \bibfield  {author} {\bibinfo {author} {\bibfnamefont {D.~S.}\ \bibnamefont
  {Chow}}, \bibinfo {author} {\bibfnamefont {F.}~\bibnamefont {Zamborszky}},
  \bibinfo {author} {\bibfnamefont {B.}~\bibnamefont {Alavi}}, \bibinfo
  {author} {\bibfnamefont {D.~J.}\ \bibnamefont {Tantillo}}, \bibinfo {author}
  {\bibfnamefont {A.}~\bibnamefont {Baur}}, \bibinfo {author} {\bibfnamefont
  {C.~A.}\ \bibnamefont {Merlic}},\ and\ \bibinfo {author} {\bibfnamefont
  {S.~E.}\ \bibnamefont {Brown}},\ }\bibfield  {title} {\bibinfo {title}
  {{Charge Ordering in the TMTTF Family of Molecular Conductors}},\ }\href
  {https://doi.org/10.1103/PhysRevLett.85.1698} {\bibfield  {journal} {\bibinfo
   {journal} {Phys. Rev. Lett.}\ }\textbf {\bibinfo {volume} {85}},\ \bibinfo
  {pages} {1698} (\bibinfo {year} {2000})}\BibitemShut {NoStop}%
\bibitem [{\citenamefont {Kitou}\ \emph {et~al.}(2021)\citenamefont {Kitou},
  \citenamefont {Zhang}, \citenamefont {Nakamura},\ and\ \citenamefont
  {Sawa}}]{Kito_2021}%
  \BibitemOpen
  \bibfield  {author} {\bibinfo {author} {\bibfnamefont {S.}~\bibnamefont
  {Kitou}}, \bibinfo {author} {\bibfnamefont {L.}~\bibnamefont {Zhang}},
  \bibinfo {author} {\bibfnamefont {T.}~\bibnamefont {Nakamura}},\ and\
  \bibinfo {author} {\bibfnamefont {H.}~\bibnamefont {Sawa}},\ }\bibfield
  {title} {\bibinfo {title} {{{Complex changes in structural parameters hidden
  in the universal phase diagram of the quasi-one-dimensional organic
  conductors ${(\mathrm{TMTTF})}_{2}X$ ($X=\mathrm{Nb}{\mathrm{F}}_{6},
  \mathrm{As}{\mathrm{F}}_{6}, \mathrm{P}{\mathrm{F}}_{6}$, and Br)}}},\ }\href
  {https://doi.org/10.1103/PhysRevB.103.184112} {\bibfield  {journal} {\bibinfo
   {journal} {Phys. Rev. B}\ }\textbf {\bibinfo {volume} {103}},\ \bibinfo
  {pages} {184112} (\bibinfo {year} {2021})}\BibitemShut {NoStop}%
\bibitem [{\citenamefont {Seo}\ and\ \citenamefont
  {Fukuyama}(1997)}]{Seo_1997}%
  \BibitemOpen
  \bibfield  {author} {\bibinfo {author} {\bibfnamefont {H.}~\bibnamefont
  {Seo}}\ and\ \bibinfo {author} {\bibfnamefont {H.}~\bibnamefont {Fukuyama}},\
  }\bibfield  {title} {\bibinfo {title} {{Antiferromagnetic Phases of
  One-Dimensional Quarter-Filled Organic Conductors}},\ }\href
  {https://doi.org/10.1143/JPSJ.66.1249} {\bibfield  {journal} {\bibinfo
  {journal} {J. Phys. Soc. Jpn.}\ }\textbf {\bibinfo {volume} {66}},\ \bibinfo
  {pages} {1249} (\bibinfo {year} {1997})}\BibitemShut {NoStop}%
\bibitem [{\citenamefont {Seo}(2000)}]{Seo_2000}%
  \BibitemOpen
  \bibfield  {author} {\bibinfo {author} {\bibfnamefont {H.}~\bibnamefont
  {Seo}},\ }\bibfield  {title} {\bibinfo {title} {{Charge Ordering in Organic
  ET Compounds}},\ }\href {https://doi.org/10.1143/JPSJ.69.805} {\bibfield
  {journal} {\bibinfo  {journal} {J. Phys. Soc. Jpn.}\ }\textbf {\bibinfo
  {volume} {69}},\ \bibinfo {pages} {805} (\bibinfo {year} {2000})}\BibitemShut
  {NoStop}%
\bibitem [{\citenamefont {Powell}\ and\ \citenamefont
  {McKenzie}(2006)}]{Powell_review2006}%
  \BibitemOpen
  \bibfield  {author} {\bibinfo {author} {\bibfnamefont {B.~J.}\ \bibnamefont
  {Powell}}\ and\ \bibinfo {author} {\bibfnamefont {R.~H.}\ \bibnamefont
  {McKenzie}},\ }\bibfield  {title} {\bibinfo {title} {{Strong electronic
  correlations in superconducting organic charge transfer salts}},\ }\href
  {https://doi.org/10.1088/0953-8984/18/45/r03} {\bibfield  {journal} {\bibinfo
   {journal} {J. Phys.: Cond. Matter.}\ }\textbf {\bibinfo {volume} {18}},\
  \bibinfo {pages} {R827} (\bibinfo {year} {2006})}\BibitemShut {NoStop}%
\bibitem [{\citenamefont {Zamborszky}\ \emph {et~al.}(2002)\citenamefont
  {Zamborszky}, \citenamefont {Yu}, \citenamefont {Raas}, \citenamefont
  {Brown}, \citenamefont {Alavi}, \citenamefont {Merlic},\ and\ \citenamefont
  {Baur}}]{Zamborszky_2002}%
  \BibitemOpen
  \bibfield  {author} {\bibinfo {author} {\bibfnamefont {F.}~\bibnamefont
  {Zamborszky}}, \bibinfo {author} {\bibfnamefont {W.}~\bibnamefont {Yu}},
  \bibinfo {author} {\bibfnamefont {W.}~\bibnamefont {Raas}}, \bibinfo {author}
  {\bibfnamefont {S.~E.}\ \bibnamefont {Brown}}, \bibinfo {author}
  {\bibfnamefont {B.}~\bibnamefont {Alavi}}, \bibinfo {author} {\bibfnamefont
  {C.~A.}\ \bibnamefont {Merlic}},\ and\ \bibinfo {author} {\bibfnamefont
  {A.}~\bibnamefont {Baur}},\ }\bibfield  {title} {\bibinfo {title}
  {{Competition and coexistence of bond and charge orders in
  $(\mathrm{TMTTF}{)}_{2}{\mathrm{AsF}}_{6}$}},\ }\href
  {https://doi.org/10.1103/PhysRevB.66.081103} {\bibfield  {journal} {\bibinfo
  {journal} {Phys. Rev. B}\ }\textbf {\bibinfo {volume} {66}},\ \bibinfo
  {pages} {081103} (\bibinfo {year} {2002})}\BibitemShut {NoStop}%
\bibitem [{\citenamefont {Iwase}\ \emph {et~al.}(2011)\citenamefont {Iwase},
  \citenamefont {Sugiura}, \citenamefont {Furukawa},\ and\ \citenamefont
  {Nakamura}}]{Iwase_2011}%
  \BibitemOpen
  \bibfield  {author} {\bibinfo {author} {\bibfnamefont {F.}~\bibnamefont
  {Iwase}}, \bibinfo {author} {\bibfnamefont {K.}~\bibnamefont {Sugiura}},
  \bibinfo {author} {\bibfnamefont {K.}~\bibnamefont {Furukawa}},\ and\
  \bibinfo {author} {\bibfnamefont {T.}~\bibnamefont {Nakamura}},\ }\bibfield
  {title} {\bibinfo {title} {{${}^{13}$C NMR study of the magnetic properties
  of the quasi-one-dimensional conductor (TMTTF)${}_{2}$SbF${}_{6}$}},\ }\href
  {https://doi.org/10.1103/PhysRevB.84.115140} {\bibfield  {journal} {\bibinfo
  {journal} {Phys. Rev. B}\ }\textbf {\bibinfo {volume} {84}},\ \bibinfo
  {pages} {115140} (\bibinfo {year} {2011})}\BibitemShut {NoStop}%
\bibitem [{\citenamefont {Dressel}\ \emph {et~al.}(2012)\citenamefont
  {Dressel}, \citenamefont {Dumm}, \citenamefont {Knoblauch},\ and\
  \citenamefont {Masino}}]{Dressel_2012}%
  \BibitemOpen
  \bibfield  {author} {\bibinfo {author} {\bibfnamefont {M.}~\bibnamefont
  {Dressel}}, \bibinfo {author} {\bibfnamefont {M.}~\bibnamefont {Dumm}},
  \bibinfo {author} {\bibfnamefont {T.}~\bibnamefont {Knoblauch}},\ and\
  \bibinfo {author} {\bibfnamefont {M.}~\bibnamefont {Masino}},\ }\bibfield
  {title} {\bibinfo {title} {{Comprehensive Optical Investigations of Charge
  Order in Organic Chain Compounds (TMTTF)$_2$X}},\ }\href
  {https://doi.org/10.3390/cryst2020528} {\bibfield  {journal} {\bibinfo
  {journal} {Crystals}\ }\textbf {\bibinfo {volume} {2}},\ \bibinfo {pages}
  {528} (\bibinfo {year} {2012})}\BibitemShut {NoStop}%
\bibitem [{\citenamefont {Kuwabara}\ \emph {et~al.}(2003)\citenamefont
  {Kuwabara}, \citenamefont {Seo},\ and\ \citenamefont
  {Ogata}}]{Kuwabara_2003}%
  \BibitemOpen
  \bibfield  {author} {\bibinfo {author} {\bibfnamefont {M.}~\bibnamefont
  {Kuwabara}}, \bibinfo {author} {\bibfnamefont {H.}~\bibnamefont {Seo}},\ and\
  \bibinfo {author} {\bibfnamefont {M.}~\bibnamefont {Ogata}},\ }\bibfield
  {title} {\bibinfo {title} {{Coexistence of Charge Order and Spin--Peierls
  Lattice Distortion in One-Dimensional Organic Conductors}},\ }\href
  {https://doi.org/10.1143/JPSJ.72.225} {\bibfield  {journal} {\bibinfo
  {journal} {J. Phys. Soc. Jpn.}\ }\textbf {\bibinfo {volume} {72}},\ \bibinfo
  {pages} {225} (\bibinfo {year} {2003})}\BibitemShut {NoStop}%
\bibitem [{\citenamefont {Sugiura}\ \emph {et~al.}(2005)\citenamefont
  {Sugiura}, \citenamefont {Tsuchiizu},\ and\ \citenamefont
  {Suzumura}}]{Sugiura_2005}%
  \BibitemOpen
  \bibfield  {author} {\bibinfo {author} {\bibfnamefont {M.}~\bibnamefont
  {Sugiura}}, \bibinfo {author} {\bibfnamefont {M.}~\bibnamefont {Tsuchiizu}},\
  and\ \bibinfo {author} {\bibfnamefont {Y.}~\bibnamefont {Suzumura}},\
  }\bibfield  {title} {\bibinfo {title} {{Spin-Peierls Transition Temperature
  in Quarter-Filled Organic Conductors}},\ }\href
  {https://doi.org/10.1143/JPSJ.74.983} {\bibfield  {journal} {\bibinfo
  {journal} {J. Phys. Soc. Jpn.}\ }\textbf {\bibinfo {volume} {74}},\ \bibinfo
  {pages} {983} (\bibinfo {year} {2005})}\BibitemShut {NoStop}%
\bibitem [{\citenamefont {Yoshimi}\ \emph
  {et~al.}(2012{\natexlab{a}})\citenamefont {Yoshimi}, \citenamefont {Seo},
  \citenamefont {Ishibashi},\ and\ \citenamefont {Brown}}]{Yoshimi_2012}%
  \BibitemOpen
  \bibfield  {author} {\bibinfo {author} {\bibfnamefont {K.}~\bibnamefont
  {Yoshimi}}, \bibinfo {author} {\bibfnamefont {H.}~\bibnamefont {Seo}},
  \bibinfo {author} {\bibfnamefont {S.}~\bibnamefont {Ishibashi}},\ and\
  \bibinfo {author} {\bibfnamefont {S.~E.}\ \bibnamefont {Brown}},\ }\bibfield
  {title} {\bibinfo {title} {{Tuning the Magnetic Dimensionality by Charge
  Ordering in the Molecular TMTTF Salts}},\ }\href
  {https://doi.org/10.1103/PhysRevLett.108.096402} {\bibfield  {journal}
  {\bibinfo  {journal} {Phys. Rev. Lett.}\ }\textbf {\bibinfo {volume} {108}},\
  \bibinfo {pages} {096402} (\bibinfo {year} {2012}{\natexlab{a}})}\BibitemShut
  {NoStop}%
\bibitem [{\citenamefont {Clay}\ and\ \citenamefont
  {Mazumdar}(2019)}]{Clay_review}%
  \BibitemOpen
  \bibfield  {author} {\bibinfo {author} {\bibfnamefont {R.~T.}\ \bibnamefont
  {Clay}}\ and\ \bibinfo {author} {\bibfnamefont {S.}~\bibnamefont
  {Mazumdar}},\ }\bibfield  {title} {\bibinfo {title} {{From charge- and
  spin-ordering to superconductivity in the organic charge-transfer solids}},\
  }\href {https://doi.org/https://doi.org/10.1016/j.physrep.2018.10.006}
  {\bibfield  {journal} {\bibinfo  {journal} {Physics Reports}\ }\textbf
  {\bibinfo {volume} {788}},\ \bibinfo {pages} {1} (\bibinfo {year}
  {2019})}\BibitemShut {NoStop}%
\bibitem [{\citenamefont {Yoshioka}\ \emph {et~al.}(2012)\citenamefont
  {Yoshioka}, \citenamefont {Otsuka},\ and\ \citenamefont
  {Seo}}]{Yoshioka_crystals_review}%
  \BibitemOpen
  \bibfield  {author} {\bibinfo {author} {\bibfnamefont {H.}~\bibnamefont
  {Yoshioka}}, \bibinfo {author} {\bibfnamefont {Y.}~\bibnamefont {Otsuka}},\
  and\ \bibinfo {author} {\bibfnamefont {H.}~\bibnamefont {Seo}},\ }\bibfield
  {title} {\bibinfo {title} {{Theoretical Studies on Phase Transitions in
  Quasi-One-Dimensional Molecular Conductors}},\ }\href
  {https://doi.org/10.3390/cryst2030996} {\bibfield  {journal} {\bibinfo
  {journal} {Crystals}\ }\textbf {\bibinfo {volume} {2}},\ \bibinfo {pages}
  {996} (\bibinfo {year} {2012})}\BibitemShut {NoStop}%
\bibitem [{sup()}]{supplement}%
  \BibitemOpen
  \href@noop {} {}\bibinfo {howpublished} {See Supplemental Material
  \url{URL_will_be_inserted_by_publisher} for details of experimental data and
  microscopic parameters in $ab$ $initio$ Hamiltonians, and the explanation of
  the screened Coulomb interaction trends and their effects on the electronic
  state, which includes Refs.~\cite{Thorup:a20141, doi:10.1021/ja00414a007,
  doi:10.1080/15421400500377511, datarepo}}\BibitemShut {NoStop}%
\bibitem [{\citenamefont {Momma}\ and\ \citenamefont {Izumi}(2011)}]{VESTA}%
  \BibitemOpen
  \bibfield  {author} {\bibinfo {author} {\bibfnamefont {K.}~\bibnamefont
  {Momma}}\ and\ \bibinfo {author} {\bibfnamefont {F.}~\bibnamefont {Izumi}},\
  }\bibfield  {title} {\bibinfo {title} {Vesta 3 for three-dimensional
  visualization of crystal, volumetric and morphology data},\ }\href
  {https://doi.org/https://doi.org/10.1107/S0021889811038970} {\bibfield
  {journal} {\bibinfo  {journal} {J. Appl. Crystallogr.}\ }\textbf {\bibinfo
  {volume} {44}},\ \bibinfo {pages} {1272} (\bibinfo {year}
  {2011})}\BibitemShut {NoStop}%
\bibitem [{\citenamefont {Dumm}\ \emph {et~al.}(2000)\citenamefont {Dumm},
  \citenamefont {Loidl}, \citenamefont {Fravel}, \citenamefont {Starkey},
  \citenamefont {Montgomery},\ and\ \citenamefont {Dressel}}]{PhysRevB.61.511}%
  \BibitemOpen
  \bibfield  {author} {\bibinfo {author} {\bibfnamefont {M.}~\bibnamefont
  {Dumm}}, \bibinfo {author} {\bibfnamefont {A.}~\bibnamefont {Loidl}},
  \bibinfo {author} {\bibfnamefont {B.~W.}\ \bibnamefont {Fravel}}, \bibinfo
  {author} {\bibfnamefont {K.~P.}\ \bibnamefont {Starkey}}, \bibinfo {author}
  {\bibfnamefont {L.~K.}\ \bibnamefont {Montgomery}},\ and\ \bibinfo {author}
  {\bibfnamefont {M.}~\bibnamefont {Dressel}},\ }\bibfield  {title} {\bibinfo
  {title} {{Electron spin resonance studies on the organic linear-chain
  compounds $(\mathrm{TMT}C\mathrm{F}{)}_{2}X (C=\mathrm{S},\mathrm{Se};$
  $X={\mathrm{PF}}_{6},{\mathrm{AsF}}_{6},{\mathrm{ClO}}_{4},\mathrm{Br})$}},\
  }\href {https://doi.org/10.1103/PhysRevB.61.511} {\bibfield  {journal}
  {\bibinfo  {journal} {Phys. Rev. B}\ }\textbf {\bibinfo {volume} {61}},\
  \bibinfo {pages} {511} (\bibinfo {year} {2000})}\BibitemShut {NoStop}%
\bibitem [{\citenamefont {Sakata}\ \emph
  {et~al.}(2006{\natexlab{a}})\citenamefont {Sakata}, \citenamefont {Yoshida},
  \citenamefont {Maesato}, \citenamefont {Saito}, \citenamefont {Matsumoto},\
  and\ \citenamefont {Hagiwara}}]{Sakata2006}%
  \BibitemOpen
  \bibfield  {author} {\bibinfo {author} {\bibfnamefont {M.}~\bibnamefont
  {Sakata}}, \bibinfo {author} {\bibfnamefont {Y.}~\bibnamefont {Yoshida}},
  \bibinfo {author} {\bibfnamefont {M.}~\bibnamefont {Maesato}}, \bibinfo
  {author} {\bibfnamefont {G.}~\bibnamefont {Saito}}, \bibinfo {author}
  {\bibfnamefont {K.}~\bibnamefont {Matsumoto}},\ and\ \bibinfo {author}
  {\bibfnamefont {R.}~\bibnamefont {Hagiwara}},\ }\bibfield  {title} {\bibinfo
  {title} {{Preparation of superconducting (TMTSF)$_2$NbF$_6$ by
  electrooxidation of TMTSF using ionic liquid as electrolyte}},\ }\href
  {https://doi.org/10.1080/15421400500377511} {\bibfield  {journal} {\bibinfo
  {journal} {Mol. Cryst. Liq. Cryst.}\ }\textbf {\bibinfo {volume} {452}},\
  \bibinfo {pages} {103} (\bibinfo {year} {2006}{\natexlab{a}})}\BibitemShut
  {NoStop}%
\bibitem [{\citenamefont {Rohwer}\ \emph {et~al.}(2020)\citenamefont {Rohwer},
  \citenamefont {Dressel},\ and\ \citenamefont {Nakamura}}]{cryst10121085}%
  \BibitemOpen
  \bibfield  {author} {\bibinfo {author} {\bibfnamefont {A.}~\bibnamefont
  {Rohwer}}, \bibinfo {author} {\bibfnamefont {M.}~\bibnamefont {Dressel}},\
  and\ \bibinfo {author} {\bibfnamefont {T.}~\bibnamefont {Nakamura}},\
  }\bibfield  {title} {\bibinfo {title} {{Deuteration Effects on the Transport
  Properties of (TMTTF)$_{2}$X Salts}},\ }\href
  {https://doi.org/10.3390/cryst10121085} {\bibfield  {journal} {\bibinfo
  {journal} {Crystals}\ }\textbf {\bibinfo {volume} {10}},\ \bibinfo {pages}
  {1085} (\bibinfo {year} {2020})}\BibitemShut {NoStop}%
\bibitem [{\citenamefont {K\"ohler}\ \emph {et~al.}(2011)\citenamefont
  {K\"ohler}, \citenamefont {Rose}, \citenamefont {Dumm}, \citenamefont
  {Untereiner},\ and\ \citenamefont {Dressel}}]{PhysRevB.84.035124}%
  \BibitemOpen
  \bibfield  {author} {\bibinfo {author} {\bibfnamefont {B.}~\bibnamefont
  {K\"ohler}}, \bibinfo {author} {\bibfnamefont {E.}~\bibnamefont {Rose}},
  \bibinfo {author} {\bibfnamefont {M.}~\bibnamefont {Dumm}}, \bibinfo {author}
  {\bibfnamefont {G.}~\bibnamefont {Untereiner}},\ and\ \bibinfo {author}
  {\bibfnamefont {M.}~\bibnamefont {Dressel}},\ }\bibfield  {title} {\bibinfo
  {title} {{Comprehensive transport study of anisotropy and ordering phenomena
  in quasi-one-dimensional (TMTTF)$_2$$X$ salts ($X$= PF$_6$, AsF$_6$, SbF$_6$,
  BF$_4$, ClO$_4$, ReO$_4$)}},\ }\href
  {https://doi.org/10.1103/PhysRevB.84.035124} {\bibfield  {journal} {\bibinfo
  {journal} {Phys. Rev. B}\ }\textbf {\bibinfo {volume} {84}},\ \bibinfo
  {pages} {035124} (\bibinfo {year} {2011})}\BibitemShut {NoStop}%
\bibitem [{\citenamefont {{Y. Nogami}}\ \emph {et~al.}(2005)\citenamefont {{Y.
  Nogami}}, \citenamefont {{T. Ito}}, \citenamefont {{K. Yamamoto}},
  \citenamefont {{N. Irie}}, \citenamefont {{S. Horita}}, \citenamefont {{T.
  Kambe}}, \citenamefont {{N. Nagao}}, \citenamefont {{K. Oshima}},
  \citenamefont {{N. Ikeda}},\ and\ \citenamefont {{T.
  Nakamura}}}]{Nogami2005}%
  \BibitemOpen
  \bibfield  {author} {\bibinfo {author} {\bibnamefont {{Y. Nogami}}}, \bibinfo
  {author} {\bibnamefont {{T. Ito}}}, \bibinfo {author} {\bibnamefont {{K.
  Yamamoto}}}, \bibinfo {author} {\bibnamefont {{N. Irie}}}, \bibinfo {author}
  {\bibnamefont {{S. Horita}}}, \bibinfo {author} {\bibnamefont {{T. Kambe}}},
  \bibinfo {author} {\bibnamefont {{N. Nagao}}}, \bibinfo {author}
  {\bibnamefont {{K. Oshima}}}, \bibinfo {author} {\bibnamefont {{N. Ikeda}}},\
  and\ \bibinfo {author} {\bibnamefont {{T. Nakamura}}},\ }\bibfield  {title}
  {\bibinfo {title} {{X-ray structural study of charge and anion orderings of
  TMTTF salts}},\ }\href {https://doi.org/10.1051/jp4:2005131008} {\bibfield
  {journal} {\bibinfo  {journal} {J. Phys. IV France}\ }\textbf {\bibinfo
  {volume} {131}},\ \bibinfo {pages} {39} (\bibinfo {year} {2005})}\BibitemShut
  {NoStop}%
\bibitem [{\citenamefont {Hohenberg}\ and\ \citenamefont
  {Kohn}(1964)}]{H-K_1964}%
  \BibitemOpen
  \bibfield  {author} {\bibinfo {author} {\bibfnamefont {P.}~\bibnamefont
  {Hohenberg}}\ and\ \bibinfo {author} {\bibfnamefont {W.}~\bibnamefont
  {Kohn}},\ }\bibfield  {title} {\bibinfo {title} {Inhomogeneous electron
  gas},\ }\href {https://doi.org/10.1103/PhysRev.136.B864} {\bibfield
  {journal} {\bibinfo  {journal} {Phys. Rev.}\ }\textbf {\bibinfo {volume}
  {136}},\ \bibinfo {pages} {B864} (\bibinfo {year} {1964})}\BibitemShut
  {NoStop}%
\bibitem [{\citenamefont {Kohn}\ and\ \citenamefont {Sham}(1965)}]{Kohn_Sham}%
  \BibitemOpen
  \bibfield  {author} {\bibinfo {author} {\bibfnamefont {W.}~\bibnamefont
  {Kohn}}\ and\ \bibinfo {author} {\bibfnamefont {L.~J.}\ \bibnamefont
  {Sham}},\ }\bibfield  {title} {\bibinfo {title} {{Self-Consistent Equations
  Including Exchange and Correlation Effects}},\ }\href
  {https://doi.org/10.1103/PhysRev.140.A1133} {\bibfield  {journal} {\bibinfo
  {journal} {Phys. Rev.}\ }\textbf {\bibinfo {volume} {140}},\ \bibinfo {pages}
  {A1133} (\bibinfo {year} {1965})}\BibitemShut {NoStop}%
\bibitem [{\citenamefont {Tahara}\ and\ \citenamefont
  {Imada}(2008)}]{Tahara_JPSJ2008}%
  \BibitemOpen
  \bibfield  {author} {\bibinfo {author} {\bibfnamefont {D.}~\bibnamefont
  {Tahara}}\ and\ \bibinfo {author} {\bibfnamefont {M.}~\bibnamefont {Imada}},\
  }\bibfield  {title} {\bibinfo {title} {{Variational Monte Carlo Method
  Combined with Quantum-Number Projection and Multi-Variable Optimization}},\
  }\href {https://doi.org/https://doi.org/10.1143/JPSJ.77.114701} {\bibfield
  {journal} {\bibinfo  {journal} {J. Phys. Soc. Jpn.}\ }\textbf {\bibinfo
  {volume} {77}},\ \bibinfo {pages} {114701} (\bibinfo {year}
  {2008})}\BibitemShut {NoStop}%
\bibitem [{\citenamefont {Misawa}\ \emph {et~al.}(2019)\citenamefont {Misawa},
  \citenamefont {Morita}, \citenamefont {Yoshimi}, \citenamefont {Kawamura},
  \citenamefont {Motoyama}, \citenamefont {Ido}, \citenamefont {Ohgoe},
  \citenamefont {Imada},\ and\ \citenamefont {Kato}}]{misawa_CPC2019}%
  \BibitemOpen
  \bibfield  {author} {\bibinfo {author} {\bibfnamefont {T.}~\bibnamefont
  {Misawa}}, \bibinfo {author} {\bibfnamefont {S.}~\bibnamefont {Morita}},
  \bibinfo {author} {\bibfnamefont {K.}~\bibnamefont {Yoshimi}}, \bibinfo
  {author} {\bibfnamefont {M.}~\bibnamefont {Kawamura}}, \bibinfo {author}
  {\bibfnamefont {Y.}~\bibnamefont {Motoyama}}, \bibinfo {author}
  {\bibfnamefont {K.}~\bibnamefont {Ido}}, \bibinfo {author} {\bibfnamefont
  {T.}~\bibnamefont {Ohgoe}}, \bibinfo {author} {\bibfnamefont
  {M.}~\bibnamefont {Imada}},\ and\ \bibinfo {author} {\bibfnamefont
  {T.}~\bibnamefont {Kato}},\ }\bibfield  {title} {\bibinfo {title}
  {{mVMC---Open-source software for many-variable variational Monte Carlo
  method}},\ }\href {https://doi.org/https://doi.org/10.1016/j.cpc.2018.08.014}
  {\bibfield  {journal} {\bibinfo  {journal} {Comput. Phys. Commun.}\ }\textbf
  {\bibinfo {volume} {235}},\ \bibinfo {pages} {447} (\bibinfo {year}
  {2019})}\BibitemShut {NoStop}%
\bibitem [{mVM()}]{mVMC}%
  \BibitemOpen
  \href@noop {} {}\bibinfo {note}
  {~https://www.pasums.issp.u-tokyo.ac.jp/mvmc/en/}\BibitemShut {NoStop}%
\bibitem [{\citenamefont {Giannozzi}\ \emph {et~al.}(2017)\citenamefont
  {Giannozzi}, \citenamefont {Andreussi}, \citenamefont {Brumme}, \citenamefont
  {Bunau}, \citenamefont {Nardelli}, \citenamefont {Calandra}, \citenamefont
  {Car}, \citenamefont {Cavazzoni}, \citenamefont {Ceresoli}, \citenamefont
  {Cococcioni} \emph {et~al.}}]{QE}%
  \BibitemOpen
  \bibfield  {author} {\bibinfo {author} {\bibfnamefont {P.}~\bibnamefont
  {Giannozzi}}, \bibinfo {author} {\bibfnamefont {O.}~\bibnamefont
  {Andreussi}}, \bibinfo {author} {\bibfnamefont {T.}~\bibnamefont {Brumme}},
  \bibinfo {author} {\bibfnamefont {O.}~\bibnamefont {Bunau}}, \bibinfo
  {author} {\bibfnamefont {M.~B.}\ \bibnamefont {Nardelli}}, \bibinfo {author}
  {\bibfnamefont {M.}~\bibnamefont {Calandra}}, \bibinfo {author}
  {\bibfnamefont {R.}~\bibnamefont {Car}}, \bibinfo {author} {\bibfnamefont
  {C.}~\bibnamefont {Cavazzoni}}, \bibinfo {author} {\bibfnamefont
  {D.}~\bibnamefont {Ceresoli}}, \bibinfo {author} {\bibfnamefont
  {M.}~\bibnamefont {Cococcioni}}, \emph {et~al.},\ }\bibfield  {title}
  {\bibinfo {title} {{Advanced capabilities for materials modelling with
  Quantum ESPRESSO}},\ }\href {https://doi.org/10.1088/1361-648X/aa8f79}
  {\bibfield  {journal} {\bibinfo  {journal} {J. of Phys.: Cond. Matt.}\
  }\textbf {\bibinfo {volume} {29}},\ \bibinfo {pages} {465901} (\bibinfo
  {year} {2017})}\BibitemShut {NoStop}%
\bibitem [{\citenamefont {Hamann}(2013)}]{Hamann_ONCV2013}%
  \BibitemOpen
  \bibfield  {author} {\bibinfo {author} {\bibfnamefont {D.~R.}\ \bibnamefont
  {Hamann}},\ }\bibfield  {title} {\bibinfo {title} {{Optimized norm-conserving
  Vanderbilt pseudopotentials}},\ }\href
  {https://doi.org/10.1103/PhysRevB.88.085117} {\bibfield  {journal} {\bibinfo
  {journal} {Phys. Rev. B}\ }\textbf {\bibinfo {volume} {88}},\ \bibinfo
  {pages} {085117} (\bibinfo {year} {2013})}\BibitemShut {NoStop}%
\bibitem [{\citenamefont {Schlipf}\ and\ \citenamefont
  {Gygi}(2015)}]{Schlipf_CPC2015}%
  \BibitemOpen
  \bibfield  {author} {\bibinfo {author} {\bibfnamefont {M.}~\bibnamefont
  {Schlipf}}\ and\ \bibinfo {author} {\bibfnamefont {F.}~\bibnamefont {Gygi}},\
  }\bibfield  {title} {\bibinfo {title} {{Optimization algorithm for the
  generation of ONCV pseudopotentials}},\ }\href
  {https://doi.org/https://doi.org/10.1016/j.cpc.2015.05.011} {\bibfield
  {journal} {\bibinfo  {journal} {Comput. Phys. Commun.}\ }\textbf {\bibinfo
  {volume} {196}},\ \bibinfo {pages} {36} (\bibinfo {year} {2015})}\BibitemShut
  {NoStop}%
\bibitem [{\citenamefont {Perdew}\ \emph {et~al.}(1996)\citenamefont {Perdew},
  \citenamefont {Burke},\ and\ \citenamefont {Ernzerhof}}]{GGA_PBE}%
  \BibitemOpen
  \bibfield  {author} {\bibinfo {author} {\bibfnamefont {J.~P.}\ \bibnamefont
  {Perdew}}, \bibinfo {author} {\bibfnamefont {K.}~\bibnamefont {Burke}},\ and\
  \bibinfo {author} {\bibfnamefont {M.}~\bibnamefont {Ernzerhof}},\ }\bibfield
  {title} {\bibinfo {title} {Generalized gradient approximation made simple},\
  }\href {https://doi.org/10.1103/PhysRevLett.77.3865} {\bibfield  {journal}
  {\bibinfo  {journal} {Phys. Rev. Lett.}\ }\textbf {\bibinfo {volume} {77}},\
  \bibinfo {pages} {3865} (\bibinfo {year} {1996})}\BibitemShut {NoStop}%
\bibitem [{\citenamefont {Ishibashi}(2009)}]{Ishibashi-2009}%
  \BibitemOpen
  \bibfield  {author} {\bibinfo {author} {\bibfnamefont {S.}~\bibnamefont
  {Ishibashi}},\ }\bibfield  {title} {\bibinfo {title} {First-principles
  electronic-band calculations on organic conductors},\ }\href
  {https://doi.org/10.1088/1468-6996/10/2/024311} {\bibfield  {journal}
  {\bibinfo  {journal} {Science and Technology of Advanced Materials}\ }\textbf
  {\bibinfo {volume} {10}},\ \bibinfo {pages} {024311} (\bibinfo {year}
  {2009})},\ \Eprint
  {https://arxiv.org/abs/https://doi.org/10.1088/1468-6996/10/2/024311}
  {https://doi.org/10.1088/1468-6996/10/2/024311} \BibitemShut {NoStop}%
\bibitem [{\citenamefont {Aryasetiawan}\ \emph {et~al.}(2004)\citenamefont
  {Aryasetiawan}, \citenamefont {Imada}, \citenamefont {Georges}, \citenamefont
  {Kotliar}, \citenamefont {Biermann},\ and\ \citenamefont
  {Lichtenstein}}]{PhysRevB.70.195104}%
  \BibitemOpen
  \bibfield  {author} {\bibinfo {author} {\bibfnamefont {F.}~\bibnamefont
  {Aryasetiawan}}, \bibinfo {author} {\bibfnamefont {M.}~\bibnamefont {Imada}},
  \bibinfo {author} {\bibfnamefont {A.}~\bibnamefont {Georges}}, \bibinfo
  {author} {\bibfnamefont {G.}~\bibnamefont {Kotliar}}, \bibinfo {author}
  {\bibfnamefont {S.}~\bibnamefont {Biermann}},\ and\ \bibinfo {author}
  {\bibfnamefont {A.~I.}\ \bibnamefont {Lichtenstein}},\ }\bibfield  {title}
  {\bibinfo {title} {Frequency-dependent local interactions and low-energy
  effective models from electronic structure calculations},\ }\href
  {https://doi.org/10.1103/PhysRevB.70.195104} {\bibfield  {journal} {\bibinfo
  {journal} {Phys. Rev. B}\ }\textbf {\bibinfo {volume} {70}},\ \bibinfo
  {pages} {195104} (\bibinfo {year} {2004})}\BibitemShut {NoStop}%
\bibitem [{\citenamefont {Imada}\ and\ \citenamefont
  {Miyake}(2010)}]{Imada_JPSJ2010}%
  \BibitemOpen
  \bibfield  {author} {\bibinfo {author} {\bibfnamefont {M.}~\bibnamefont
  {Imada}}\ and\ \bibinfo {author} {\bibfnamefont {T.}~\bibnamefont {Miyake}},\
  }\bibfield  {title} {\bibinfo {title} {Electronic structure calculation by
  first principles for strongly correlated electron systems},\ }\href
  {https://doi.org/10.1143/JPSJ.79.112001} {\bibfield  {journal} {\bibinfo
  {journal} {J. Phys. Soc. Jpn.}\ }\textbf {\bibinfo {volume} {79}},\ \bibinfo
  {pages} {112001} (\bibinfo {year} {2010})}\BibitemShut {NoStop}%
\bibitem [{\citenamefont {Shinaoka}\ \emph {et~al.}(2012)\citenamefont
  {Shinaoka}, \citenamefont {Misawa}, \citenamefont {Nakamura},\ and\
  \citenamefont {Imada}}]{Shinaoka2012}%
  \BibitemOpen
  \bibfield  {author} {\bibinfo {author} {\bibfnamefont {H.}~\bibnamefont
  {Shinaoka}}, \bibinfo {author} {\bibfnamefont {T.}~\bibnamefont {Misawa}},
  \bibinfo {author} {\bibfnamefont {K.}~\bibnamefont {Nakamura}},\ and\
  \bibinfo {author} {\bibfnamefont {M.}~\bibnamefont {Imada}},\ }\bibfield
  {title} {\bibinfo {title} {{Mott Transition and Phase Diagram of
  $\kappa$-(BEDT-TTF)2Cu(NCS)2 Studied by Two-Dimensional Model Derived from Ab
  initio Method}},\ }\href {https://doi.org/10.1143/JPSJ.81.034701} {\bibfield
  {journal} {\bibinfo  {journal} {J. Phys. Soc. Jpn}\ }\textbf {\bibinfo
  {volume} {81}},\ \bibinfo {pages} {034701} (\bibinfo {year}
  {2012})}\BibitemShut {NoStop}%
\bibitem [{\citenamefont {Misawa}\ \emph {et~al.}(2020)\citenamefont {Misawa},
  \citenamefont {Yoshimi},\ and\ \citenamefont
  {Tsumuraya}}]{PhysRevResearch.2.032072}%
  \BibitemOpen
  \bibfield  {author} {\bibinfo {author} {\bibfnamefont {T.}~\bibnamefont
  {Misawa}}, \bibinfo {author} {\bibfnamefont {K.}~\bibnamefont {Yoshimi}},\
  and\ \bibinfo {author} {\bibfnamefont {T.}~\bibnamefont {Tsumuraya}},\
  }\bibfield  {title} {\bibinfo {title} {{Electronic correlation and
  geometrical frustration in molecular solids: A systematic ab initio study of
  ${\ensuremath{\beta}}^{\ensuremath{'}}-X{[\mathrm{Pd}{(\mathrm{dmit})}_{2}]}_{2}$}},\
  }\href {https://doi.org/10.1103/PhysRevResearch.2.032072} {\bibfield
  {journal} {\bibinfo  {journal} {Phys. Rev. Research}\ }\textbf {\bibinfo
  {volume} {2}},\ \bibinfo {pages} {032072} (\bibinfo {year}
  {2020})}\BibitemShut {NoStop}%
\bibitem [{\citenamefont {Yoshimi}\ \emph {et~al.}(2021)\citenamefont
  {Yoshimi}, \citenamefont {Tsumuraya},\ and\ \citenamefont
  {Misawa}}]{PhysRevResearch.3.043224}%
  \BibitemOpen
  \bibfield  {author} {\bibinfo {author} {\bibfnamefont {K.}~\bibnamefont
  {Yoshimi}}, \bibinfo {author} {\bibfnamefont {T.}~\bibnamefont {Tsumuraya}},\
  and\ \bibinfo {author} {\bibfnamefont {T.}~\bibnamefont {Misawa}},\
  }\bibfield  {title} {\bibinfo {title} {Ab initio derivation and exact
  diagonalization analysis of low-energy effective hamiltonians for
  ${\ensuremath{\beta}}^{\ensuremath{'}}$\rm{-}$\mathrm{X}{[\mathrm{Pd}{(\mathrm{dmit})}_{2}]}_{2}$},\
  }\href {https://doi.org/10.1103/PhysRevResearch.3.043224} {\bibfield
  {journal} {\bibinfo  {journal} {Phys. Rev. Research}\ }\textbf {\bibinfo
  {volume} {3}},\ \bibinfo {pages} {043224} (\bibinfo {year}
  {2021})}\BibitemShut {NoStop}%
\bibitem [{\citenamefont {Ido}\ \emph {et~al.}(2022)\citenamefont {Ido},
  \citenamefont {Yoshimi}, \citenamefont {Misawa},\ and\ \citenamefont
  {Imada}}]{Ido2022}%
  \BibitemOpen
  \bibfield  {author} {\bibinfo {author} {\bibfnamefont {K.}~\bibnamefont
  {Ido}}, \bibinfo {author} {\bibfnamefont {K.}~\bibnamefont {Yoshimi}},
  \bibinfo {author} {\bibfnamefont {T.}~\bibnamefont {Misawa}},\ and\ \bibinfo
  {author} {\bibfnamefont {M.}~\bibnamefont {Imada}},\ }\bibfield  {title}
  {\bibinfo {title} {Unconventional dual 1d--2d quantum spin liquid revealed by
  ab initio studies on organic solids family},\ }\href
  {https://doi.org/10.1038/s41535-022-00452-8} {\bibfield  {journal} {\bibinfo
  {journal} {npj Quantum Mater.}\ }\textbf {\bibinfo {volume} {7}},\ \bibinfo
  {pages} {48} (\bibinfo {year} {2022})}\BibitemShut {NoStop}%
\bibitem [{\citenamefont {Ohki}\ \emph {et~al.}(2022)\citenamefont {Ohki},
  \citenamefont {Yoshimi},\ and\ \citenamefont
  {Kobayashi}}]{PhysRevB.105.205123}%
  \BibitemOpen
  \bibfield  {author} {\bibinfo {author} {\bibfnamefont {D.}~\bibnamefont
  {Ohki}}, \bibinfo {author} {\bibfnamefont {K.}~\bibnamefont {Yoshimi}},\ and\
  \bibinfo {author} {\bibfnamefont {A.}~\bibnamefont {Kobayashi}},\ }\bibfield
  {title} {\bibinfo {title} {{Interaction-induced quantum spin Hall insulator
  in the organic Dirac electron system
  $\ensuremath{\alpha}\text{\ensuremath{-}}{\text{(BEDT-TSeF)}}_{2}{\mathrm{I}}_{3}$}},\
  }\href {https://doi.org/10.1103/PhysRevB.105.205123} {\bibfield  {journal}
  {\bibinfo  {journal} {Phys. Rev. B}\ }\textbf {\bibinfo {volume} {105}},\
  \bibinfo {pages} {205123} (\bibinfo {year} {2022})}\BibitemShut {NoStop}%
\bibitem [{\citenamefont {Ohki}\ \emph {et~al.}(2020)\citenamefont {Ohki},
  \citenamefont {Yoshimi},\ and\ \citenamefont
  {Kobayashi}}]{PhysRevB.102.235116}%
  \BibitemOpen
  \bibfield  {author} {\bibinfo {author} {\bibfnamefont {D.}~\bibnamefont
  {Ohki}}, \bibinfo {author} {\bibfnamefont {K.}~\bibnamefont {Yoshimi}},\ and\
  \bibinfo {author} {\bibfnamefont {A.}~\bibnamefont {Kobayashi}},\ }\bibfield
  {title} {\bibinfo {title} {{Transport properties of the organic Dirac
  electron system
  $\ensuremath{\alpha}\text{\ensuremath{-}}{(\mathrm{BEDT}\text{\ensuremath{-}}\mathrm{TSeF})}_{2}{\mathrm{I}}_{3}$}},\
  }\href {https://doi.org/10.1103/PhysRevB.102.235116} {\bibfield  {journal}
  {\bibinfo  {journal} {Phys. Rev. B}\ }\textbf {\bibinfo {volume} {102}},\
  \bibinfo {pages} {235116} (\bibinfo {year} {2020})}\BibitemShut {NoStop}%
\bibitem [{\citenamefont {Nakamura}\ \emph {et~al.}(2021)\citenamefont
  {Nakamura}, \citenamefont {Yoshimoto}, \citenamefont {Nomura}, \citenamefont
  {Tadano}, \citenamefont {Kawamura}, \citenamefont {Kosugi}, \citenamefont
  {Yoshimi}, \citenamefont {Misawa},\ and\ \citenamefont {Motoyama}}]{RESPACK}%
  \BibitemOpen
  \bibfield  {author} {\bibinfo {author} {\bibfnamefont {K.}~\bibnamefont
  {Nakamura}}, \bibinfo {author} {\bibfnamefont {Y.}~\bibnamefont {Yoshimoto}},
  \bibinfo {author} {\bibfnamefont {Y.}~\bibnamefont {Nomura}}, \bibinfo
  {author} {\bibfnamefont {T.}~\bibnamefont {Tadano}}, \bibinfo {author}
  {\bibfnamefont {M.}~\bibnamefont {Kawamura}}, \bibinfo {author}
  {\bibfnamefont {T.}~\bibnamefont {Kosugi}}, \bibinfo {author} {\bibfnamefont
  {K.}~\bibnamefont {Yoshimi}}, \bibinfo {author} {\bibfnamefont
  {T.}~\bibnamefont {Misawa}},\ and\ \bibinfo {author} {\bibfnamefont
  {Y.}~\bibnamefont {Motoyama}},\ }\bibfield  {title} {\bibinfo {title}
  {{RESPACK: An ab initio tool for derivation of effective low-energy model of
  material}},\ }\href {https://doi.org/doi.org/10.1016/j.cpc.2020.107781}
  {\bibfield  {journal} {\bibinfo  {journal} {Comp. Phys. Comm.}\ }\textbf
  {\bibinfo {volume} {261}},\ \bibinfo {pages} {107781} (\bibinfo {year}
  {2021})}\BibitemShut {NoStop}%
\bibitem [{\citenamefont {Yoshimi}\ \emph
  {et~al.}(2012{\natexlab{b}})\citenamefont {Yoshimi}, \citenamefont {Seo},
  \citenamefont {Ishibashi},\ and\ \citenamefont
  {Brown}}]{Yoshimi2012_PhysicaB}%
  \BibitemOpen
  \bibfield  {author} {\bibinfo {author} {\bibfnamefont {K.}~\bibnamefont
  {Yoshimi}}, \bibinfo {author} {\bibfnamefont {H.}~\bibnamefont {Seo}},
  \bibinfo {author} {\bibfnamefont {S.}~\bibnamefont {Ishibashi}},\ and\
  \bibinfo {author} {\bibfnamefont {S.~E.}\ \bibnamefont {Brown}},\ }\bibfield
  {title} {\bibinfo {title} {{Spin frustration, charge ordering, and enhanced
  antiferromagnetism in TMTTF$_2$SbF$_6$}},\ }\href
  {https://doi.org/https://doi.org/10.1016/j.physb.2012.01.029} {\bibfield
  {journal} {\bibinfo  {journal} {Physica B: Condensed Matter}\ }\textbf
  {\bibinfo {volume} {407}},\ \bibinfo {pages} {1783} (\bibinfo {year}
  {2012}{\natexlab{b}})}\BibitemShut {NoStop}%
\bibitem [{\citenamefont {Jacko}\ \emph {et~al.}(2013)\citenamefont {Jacko},
  \citenamefont {Feldner}, \citenamefont {Rose}, \citenamefont {Lissner},
  \citenamefont {Dressel}, \citenamefont {Valent\'{\i}},\ and\ \citenamefont
  {Jeschke}}]{Jacko2013}%
  \BibitemOpen
  \bibfield  {author} {\bibinfo {author} {\bibfnamefont {A.~C.}\ \bibnamefont
  {Jacko}}, \bibinfo {author} {\bibfnamefont {H.}~\bibnamefont {Feldner}},
  \bibinfo {author} {\bibfnamefont {E.}~\bibnamefont {Rose}}, \bibinfo {author}
  {\bibfnamefont {F.}~\bibnamefont {Lissner}}, \bibinfo {author} {\bibfnamefont
  {M.}~\bibnamefont {Dressel}}, \bibinfo {author} {\bibfnamefont
  {R.}~\bibnamefont {Valent\'{\i}}},\ and\ \bibinfo {author} {\bibfnamefont
  {H.~O.}\ \bibnamefont {Jeschke}},\ }\bibfield  {title} {\bibinfo {title}
  {Electronic properties of fabre charge-transfer salts under various
  temperature and pressure conditions},\ }\href
  {https://doi.org/10.1103/PhysRevB.87.155139} {\bibfield  {journal} {\bibinfo
  {journal} {Phys. Rev. B}\ }\textbf {\bibinfo {volume} {87}},\ \bibinfo
  {pages} {155139} (\bibinfo {year} {2013})}\BibitemShut {NoStop}%
\bibitem [{\citenamefont {Gutzwiller}(1963)}]{Gutzwiller_PRL1963}%
  \BibitemOpen
  \bibfield  {author} {\bibinfo {author} {\bibfnamefont {M.~C.}\ \bibnamefont
  {Gutzwiller}},\ }\bibfield  {title} {\bibinfo {title} {{Effect of Correlation
  on the Ferromagnetism of Transition Metals}},\ }\href
  {https://doi.org/10.1103/PhysRevLett.10.159} {\bibfield  {journal} {\bibinfo
  {journal} {Phys. Rev. Lett.}\ }\textbf {\bibinfo {volume} {10}},\ \bibinfo
  {pages} {159} (\bibinfo {year} {1963})}\BibitemShut {NoStop}%
\bibitem [{\citenamefont {Jastrow}(1955)}]{Jastrow_PR1955}%
  \BibitemOpen
  \bibfield  {author} {\bibinfo {author} {\bibfnamefont {R.}~\bibnamefont
  {Jastrow}},\ }\bibfield  {title} {\bibinfo {title} {{Many-Body Problem with
  Strong Forces}},\ }\href {https://doi.org/10.1103/PhysRev.98.1479} {\bibfield
   {journal} {\bibinfo  {journal} {Phys. Rev.}\ }\textbf {\bibinfo {volume}
  {98}},\ \bibinfo {pages} {1479} (\bibinfo {year} {1955})}\BibitemShut
  {NoStop}%
\bibitem [{\citenamefont {Sorella}(2001)}]{Sorella_PRB2001}%
  \BibitemOpen
  \bibfield  {author} {\bibinfo {author} {\bibfnamefont {S.}~\bibnamefont
  {Sorella}},\ }\bibfield  {title} {\bibinfo {title} {{Generalized Lanczos
  algorithm for variational quantum Monte Carlo}},\ }\href
  {https://doi.org/10.1103/PhysRevB.64.024512} {\bibfield  {journal} {\bibinfo
  {journal} {Phys. Rev. B}\ }\textbf {\bibinfo {volume} {64}},\ \bibinfo
  {pages} {024512} (\bibinfo {year} {2001})}\BibitemShut {NoStop}%
\bibitem [{\citenamefont {Nakamura}\ \emph {et~al.}(2010)\citenamefont
  {Nakamura}, \citenamefont {Yoshimoto}, \citenamefont {Nohara},\ and\
  \citenamefont {Imada}}]{Nakamura_JPSJ2010}%
  \BibitemOpen
  \bibfield  {author} {\bibinfo {author} {\bibfnamefont {K.}~\bibnamefont
  {Nakamura}}, \bibinfo {author} {\bibfnamefont {Y.}~\bibnamefont {Yoshimoto}},
  \bibinfo {author} {\bibfnamefont {Y.}~\bibnamefont {Nohara}},\ and\ \bibinfo
  {author} {\bibfnamefont {M.}~\bibnamefont {Imada}},\ }\bibfield  {title}
  {\bibinfo {title} {{Ab initio low-dimensional physics opened up by
  dimensional downfolding: application to LaFeAsO}},\ }\href
  {https://doi.org/https://doi.org/10.1143/JPSJ.79.123708} {\bibfield
  {journal} {\bibinfo  {journal} {J. Phys. Soc. Jpn}\ }\textbf {\bibinfo
  {volume} {79}},\ \bibinfo {pages} {123708} (\bibinfo {year}
  {2010})}\BibitemShut {NoStop}%
\bibitem [{\citenamefont {Nakamura}\ \emph {et~al.}(2012)\citenamefont
  {Nakamura}, \citenamefont {Yoshimoto},\ and\ \citenamefont
  {Imada}}]{PhysRevB.86.205117}%
  \BibitemOpen
  \bibfield  {author} {\bibinfo {author} {\bibfnamefont {K.}~\bibnamefont
  {Nakamura}}, \bibinfo {author} {\bibfnamefont {Y.}~\bibnamefont
  {Yoshimoto}},\ and\ \bibinfo {author} {\bibfnamefont {M.}~\bibnamefont
  {Imada}},\ }\bibfield  {title} {\bibinfo {title} {Ab initio two-dimensional
  multiband low-energy models of
  $\rm{EtMe}_{3}\rm{Sb}$[$\rm{Pd}$(dmit)${}_{2}$]${}_{2}$ and
  $\ensuremath{\kappa}$-(\rm{BEDT-TTF})${}_{2}$\rm{Cu(NCS)}${}_{2}$ with
  comparisons to single-band models},\ }\href
  {https://doi.org/10.1103/PhysRevB.86.205117} {\bibfield  {journal} {\bibinfo
  {journal} {Phys. Rev. B}\ }\textbf {\bibinfo {volume} {86}},\ \bibinfo
  {pages} {205117} (\bibinfo {year} {2012})}\BibitemShut {NoStop}%
\bibitem [{\citenamefont {Thorup}\ \emph {et~al.}(1981)\citenamefont {Thorup},
  \citenamefont {Rindorf}, \citenamefont {Soling},\ and\ \citenamefont
  {Bechgaard}}]{Thorup:a20141}%
  \BibitemOpen
  \bibfield  {author} {\bibinfo {author} {\bibfnamefont {N.}~\bibnamefont
  {Thorup}}, \bibinfo {author} {\bibfnamefont {G.}~\bibnamefont {Rindorf}},
  \bibinfo {author} {\bibfnamefont {H.}~\bibnamefont {Soling}},\ and\ \bibinfo
  {author} {\bibfnamefont {K.}~\bibnamefont {Bechgaard}},\ }\bibfield  {title}
  {\bibinfo {title} {{The structure of
  di(2,3,6,7-tetramethyl-1,4,5,8-tetraselenafulvalenium) hexafluorophosphate,
  (TMTSF)${\sb 2}$PF${\sb 6}$, the first superconducting organic solid}},\
  }\href {https://doi.org/10.1107/S0567740881005566} {\bibfield  {journal}
  {\bibinfo  {journal} {Acta Crystallographica Section B}\ }\textbf {\bibinfo
  {volume} {37}},\ \bibinfo {pages} {1236} (\bibinfo {year}
  {1981})}\BibitemShut {NoStop}%
\bibitem [{\citenamefont {Wudl}(1981)}]{doi:10.1021/ja00414a007}%
  \BibitemOpen
  \bibfield  {author} {\bibinfo {author} {\bibfnamefont {F.}~\bibnamefont
  {Wudl}},\ }\bibfield  {title} {\bibinfo {title} {Three-dimensional structure
  of the superconductor \rm{(TMTSF)}$_2$\rm{AsF}$_6$ and the spin-charge
  separation hypothesis},\ }\href {https://doi.org/10.1021/ja00414a007}
  {\bibfield  {journal} {\bibinfo  {journal} {Journal of the American Chemical
  Society}\ }\textbf {\bibinfo {volume} {103}},\ \bibinfo {pages} {7064}
  (\bibinfo {year} {1981})},\ \Eprint
  {https://arxiv.org/abs/https://doi.org/10.1021/ja00414a007}
  {https://doi.org/10.1021/ja00414a007} \BibitemShut {NoStop}%
\bibitem [{\citenamefont {Sakata}\ \emph
  {et~al.}(2006{\natexlab{b}})\citenamefont {Sakata}, \citenamefont {Yoshida},
  \citenamefont {Maesato}, \citenamefont {Saito}, \citenamefont {Matsumoto},\
  and\ \citenamefont {Hagiwara}}]{doi:10.1080/15421400500377511}%
  \BibitemOpen
  \bibfield  {author} {\bibinfo {author} {\bibfnamefont {M.}~\bibnamefont
  {Sakata}}, \bibinfo {author} {\bibfnamefont {Y.}~\bibnamefont {Yoshida}},
  \bibinfo {author} {\bibfnamefont {M.}~\bibnamefont {Maesato}}, \bibinfo
  {author} {\bibfnamefont {G.}~\bibnamefont {Saito}}, \bibinfo {author}
  {\bibfnamefont {K.}~\bibnamefont {Matsumoto}},\ and\ \bibinfo {author}
  {\bibfnamefont {R.}~\bibnamefont {Hagiwara}},\ }\bibfield  {title} {\bibinfo
  {title} {Preparation of superconducting \rm{(TMTSF)}$_2$\rm{NbF}$_6$ by
  electrooxidation of tmtsf using ionic liquid as electrolyte},\ }\href
  {https://doi.org/10.1080/15421400500377511} {\bibfield  {journal} {\bibinfo
  {journal} {Molecular Crystals and Liquid Crystals}\ }\textbf {\bibinfo
  {volume} {452}},\ \bibinfo {pages} {103} (\bibinfo {year}
  {2006}{\natexlab{b}})},\ \Eprint
  {https://arxiv.org/abs/https://doi.org/10.1080/15421400500377511}
  {https://doi.org/10.1080/15421400500377511} \BibitemShut {NoStop}%
\bibitem [{dat()}]{datarepo}%
  \BibitemOpen
  \href@noop {} {\bibinfo {title}
  {\url{https://isspns-gitlab.issp.u-tokyo.ac.jp/k-yoshimi/tm-salts}}}\BibitemShut
  {NoStop}%
\end{thebibliography}%

\clearpage
\noindent

{\Large Supplemental Materials for
``Comprehensive $ab$ $initio$ investigation of the phase diagram of quasi-one-dimensional molecular solids"}

\section{1.~Transition temperatures}
In Table~\ref{experiment_T}, we list the experimental transition temperatures reported in the literature, which we plot in Fig.~1(a) to construct the unified phase diagram of the TM$_2${\it X} family. 
$T_{\rm DMI}$ is the temperature where the electrical resistivity takes the minimum (sometimes written as $T_{\rho}$), below which the dimer-Mott insulating state is stabilized; there is no symmetry breaking there, since it is a crossover temperature. 
Others exhibit symmetry breaking and show continuous phase transitions. 

\begin{table}[b]
\caption{List of the transition temperatures (in Kelvin) for dimer{-}Mott insulator(DMI), charge ordering (CO), {antiferromagnetic} (AF{M}), spin-density{-}wave (SDW), and spin-Peierls (SP) phases.
}
  \begin{tabular}{lcccc} \hline
     &$T_{\rm DMI}$& $T_{\rm CO}$ &$T_{\rm AF{M}/SDW}$ & $T_{\rm SP}$ \\ \hline
     {TMTSF}$_2$PF$_6$\cite{PhysRevB.61.511} &--& -- & 12 & --\\
     {TMTSF}$_2$AsF$_6$\cite{PhysRevB.61.511} &--&-- & 12 & --\\
     {TMTSF}$_2$NbF$_6$\cite{Sakata2006}&--&-- & 12 & --\\
     \hline
    {TMTTF}$_2$Br\cite{cryst10121085} &130 & 30 & 13& --\\
    {TMTTF}$_2$PF$_6$\cite{PhysRevB.84.035124} &250 & 67 & -- & 19\\
    {TMTTF}$_2$AsF$_6$\cite{PhysRevB.84.035124} &250& 102 & -- & 19\\
    {TMTTF}$_2$SbF$_6$\cite{PhysRevB.84.035124} &240 & 157 & 6 & --\\
    {TMTTF}$_2$NbF$_6$\cite{Kito_2021} & \ -- \footnote{Although the presence of DMI phase has been reported in Ref.~\cite{Kito_2021}, the value of $T_{\rm DMI}$ is not reported.} & 165 & 10 & --\\
     \hline
  \end{tabular}
\label{experiment_T}
\end{table}

\section{2.~Model parameters}
Since molecular solids have strong temperature dependencies on structure parameters due to their softness, 
the model parameters derived for structures at different temperatures can be different.
In fact, it is shown that in a Pd(dmit)$_2$ based system, as the temperature decreases, 
the amplitude of electron correlation effectively decreases~\cite{PhysRevResearch.3.043224}. 
Therefore, in this work, we performed a systematic analysis based on the structures at a fixed temperature of $200$~K for TMTTF$_2$X (X=Br, PF$_6$, AsF$_6$, and NbF$_6$) provided by Kitou {\it et al}~\cite{Kito_2021}. 
The analysis was based on room temperature structures {for TMTSF$_2$X (X=PF$_6$\cite{Thorup:a20141}, AsF$_6$\cite{doi:10.1021/ja00414a007}, and NbF$_6$\cite{doi:10.1080/15421400500377511})}. 
The conclusion that the electron correlation effect is weaker for the TMTSF salts than 
for the TMTTF salts holds, despite the difference in the temperature used between the two systems, since the amplitude of the correlations in TMTSF becomes weaker at lower temperatures.

\begin{table*}
\caption{List of the parameters obtained by the downfolding method. {The unit of unitcell volume $V_{\rm uc}$ is \AA$^3$.} The energy unit is eV. {{-} in $t_{q2}$ indicates that the absolute values are less than $10^{-3}$ eV.}}
  \begin{tabular}{lccccccccccccccc} \hline
     & $T$ & {$V_{\rm uc}$} & $t_{a1}$ & $t_{a2}$ & $t_{b}$ &$t_{q1}$ & $t_{q2}$ &{$U_{\rm bare}$}&$U$& $V_{a1}$ & $V_{a2}$   & $V_{b}$  & $V_{q1}$  & $V_{q2}$  & $V_{2a}$\\ \hline
     {TMTSF}$_2$PF$_6$ & RT   & {714.3}& {-0.209} & {-0.235} & {0.050} & {0.027} & {0.006} & {5.27}& {2.13} & {0.998} & {1.016}  & {0.645}& {0.731} & {0.635}& {0.588}\\
     {TMTSF}$_2$AsF$_6$ & RT   & {719.9} & -0.210 & -0.232 & 0.044 & 0.019 & 0.004 & {5.19}& 2.09 & 0.996 & 1.009  & 0.636 & 0.724 & 0.624 & 0.589\\
     {TMTSF}$_2$NbF$_6$ & RT  & {735.3} &-0.195 & -0.217 & 0.037& 0.019 & - & {5.21} &2.18 & 1.05 & 1.07 & 0.662  & 0.751 & 0.650 & 0.619\\
     \hline
    {TMTTF}$_2$Br & 200K  & {611.7} & -0.188 & -0.204 & 0.037 & 0.029 & 0.019 & {5.54}& 2.32 & 1.06 & 1.07 & 0.635& 0.709 & 0.647 &  0.603\\
    {TMTTF}$_2$PF$_6$ & 200K  & {661.4} & -0.173 & -0.218 & 0.03& 0.012 & 0.005 & {5.64}& 2.50 & 1.14 & 1.18 & 0.70& 0.788 & 0.684 & 0.666\\
    {TMTTF}$_2$AsF$_6$ & 200K   & {672.6} &-0.178 & -0.213 & 0.03& 0.011 & - & {5.64}& 2.50  & 1.14 & 1.18 & 0.70& 0.783 & 0.677 &  0.664\\
    {TMTTF}$_2$NbF$_6$ & 200K   & {683.0} & -0.168 & -0.209 & 0.03& 0.011 & - & {5.64}& 2.49 & 1.14 & 1.17 & 0.688& 0.767 & 0.667 & 0.66
\\ \hline
  \end{tabular}
\label{TransferUV}
\end{table*}

In Table~\ref{TransferUV}, we show the calculated values of the transfer integrals{, the bare on-site Coulomb interaction $U_{\rm bare}$, and the screened} Coulomb interactions using DFT and MLWFs. 
{The unitcell volumes $V_{\rm uc}$ obtained by the crystallographic information file are also listed.}
Due to the numerical errors, values of crystallographically equivalent bonds were slightly different. We averaged over the transfer integrals and Coulomb interactions for corresponding bonds. {We also confirmed that in the actual analysis of the low-energy effective models, meV order differences in the parameter do not change the results.} We uploaded the output files containing the transition integrals 
and Coulomb interactions obtained by RESPACK to the data repository\cite{datarepo}.

\begin{figure}[t] 
\begin{center} 
\includegraphics[width=0.5 \textwidth]{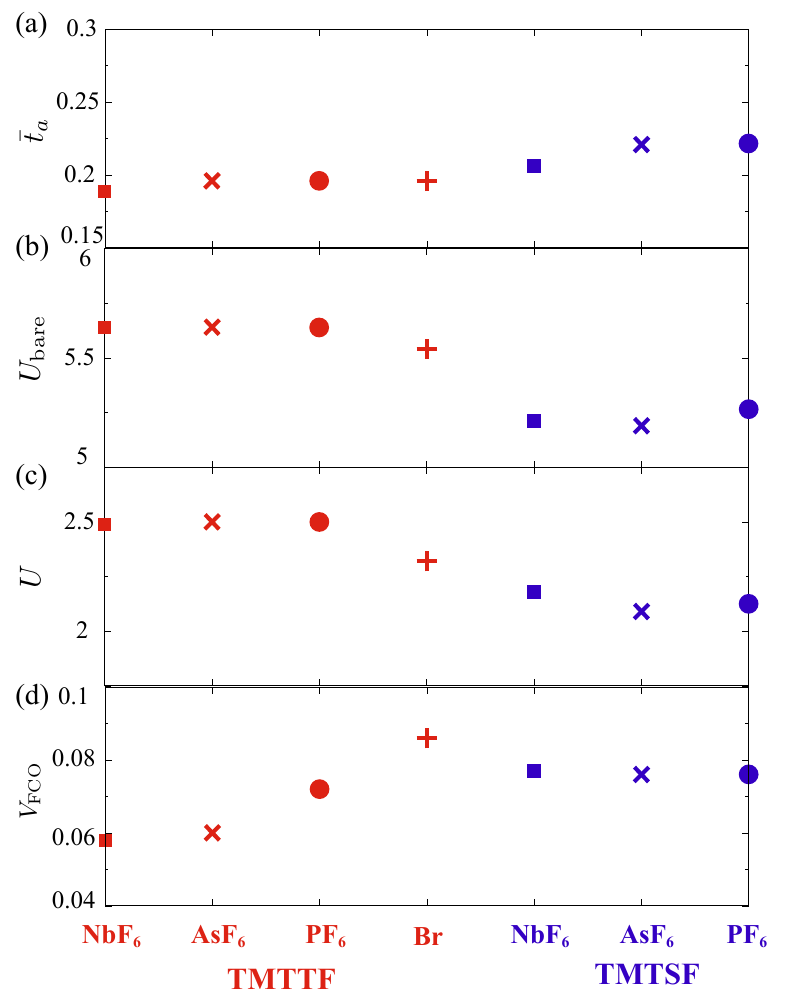}
\caption{
{(a) Mean transfer integral  between TM molecules along $a$-axis ($\bar{t}_a\equiv |(t_{a1}+t_{a2})/2|$),
(b) the {bare} on-site Coulomb interaction ($U_{\rm bare}$), (c) the on-site Coulomb interaction ($U$), and (d)  ${V_{\rm FCO}}\equiv V_{q1}+V_{q2}-2V_{b}$ for TM$_2${\it X} (TM = TMTSF, TMTTF, {\it X} = NbF$_6$, AsF$_6$, Br, and PF$_6$).}}
\label{fig:Params}
\end{center}
\end{figure}

Figure \ref{fig:Params} (a) shows mean transfer integral between TM molecules along $a$-axis ($\bar{t}_a\equiv |(t_{a1}+t_{a2})/2|$). The obtained results can be understood by the effect of chemical pressure, namely the change in the unit cell volume. When the radius of the anion is small, the unit cell volume becomes small and then the transfer integrals increase. 
Figure \ref{fig:Params} (b) shows the bare on-site Coulomb interaction $U_{\rm bare}$. Since $U_{\rm bare}$ is determined by the charge distribution on the TM molecule,  the magnitude of $U_{\rm bare}$ depends mostly on the type of TM molecule (for TMTSF molecules, $U_{\rm bare}$ is smaller than for TMTTF molecules because the electron density in the molecule expands, by the difference between Sulfur in TMTTF molecule and Selenium in TMTSF molecule). As the transfer integrals increase, in the cRPA method, the screening effect increases and thus the screened on-site Coulomb interaction $U$ decreases as shown in Fig. \ref{fig:Params} (c). As a result, the material dependence of $U/{\bar{t}_a}$ becomes pronounced  and systematically varied among the investigated members.

The intersite Coulomb interactions are crucial in stabilizing the CO state. 
Since the Coulomb interactions $V_{a1}$ and $V_{a2}$ are notably large compared to other Coulomb interactions, the ``rich-poor" charge pattern along the $a$ axis is favored. 
Thus, in order to stabilize the ferroelectric state, the classical condition is that sum of the long-range Coulomb interactions connected by $q1$ and $q2$ bonds are larger than twice of that 
connected by the $b$ bond: $V_{\rm FCO}\equiv V_{q1}+V_{q2}-2V_b > 0$.
As seen from Fig. \ref{fig:Params} (d), the intersite Coulomb interactions derived in an $ab$ $initio$ way here indeed satisfy this condition for all of the compounds. 

\end{document}